\documentclass[aps,prd,floatfix,superscriptaddress,preprintnumbers,nofootinbib,twocolumn,longbibliography]{revtex4-1}
\pdfoutput=1

\usepackage{amsmath, amssymb, amsfonts, graphicx, mathrsfs, verbatim, float, slashed, bbm, color, xcolor, xspace, enumerate, url, bbding, multirow, outlines,upgreek}
\usepackage[english]{babel}
\usepackage[colorlinks=true,linkcolor=black,citecolor=galacticcenterbubblegum]{hyperref}
\definecolor{galacticcenterbubblegum}{rgb}{0.8,0, 0.8}

\usepackage{graphicx,bm}

\newcommand{\beq}{\begin{equation}}
\newcommand{\bea}{\begin{eqnarray}}
\newcommand{\eeq}{\end{equation}}
\newcommand{\eea}{\end{eqnarray}}
\newcommand{\bal}{\begin{align}}
\newcommand{\eal}{\end{align}}

\usepackage{tikz}
\usetikzlibrary{decorations.pathmorphing,decorations.markings}
\tikzset{
photon/.style={decorate, decoration={snake,amplitude=4pt, segment length=7pt}, draw=black},
particle/.style={draw=black, postaction={decorate}, decoration={markings,mark=at position .5 with {\arrow[draw=black]{>}}}},
antiparticle/.style={draw=black, postaction={decorate}, decoration={markings,mark=at position .5 with {\arrow[draw=black]{<}}}},
gluon/.style={decorate, draw=black, decoration={coil,amplitude=3pt, segment length=4pt}},
higgs/.style={draw=black,dashed,thick },
arrow/.style={draw=black, very thick, postaction={decorate}, decoration={markings,mark=at position 1 with {\arrow[draw=black]{>}}}}
}

\usepackage{latexsym}
\usepackage{subcaption}

\newcommand{\gev}{~\mathrm{GeV}}
\newcommand{\TeV}{~\mathrm{TeV}}

\usepackage{outlines}
\usepackage{enumitem}
\setenumerate[1]{label=\Roman*.}
\setenumerate[2]{label=\Alph*.}
\setenumerate[3]{label=\roman*.}
\setenumerate[4]{label=\alph*.}

\newcommand{\Lc}{\mathcal{L}}

\newcommand{\G}{\text{G}}

\newcommand{\Z}{\mathbb{Z}}

\newcommand{\FN}{\text{FN}}
\newcommand{\T}{T_\star}
\newcommand{\Tx}{T_{\text{cf}}}
\newcommand{\TEW}{T_{\text{EW}}}
\newcommand{\TBmin}{T_{B_Y(\text{min})}}
\newcommand{\TBmax}{T_{B_Y(\text{max})}}
\newcommand{\Tsat}{T^{\text{non-CME}}_{E\cdot B}}
\newcommand{\Tmax}{T_{\text{max}}}
\newcommand{\Mpl}{M_{\text{Pl}}}

\definecolor{darklightsabergreen}{rgb}{0.0, .49, 0.06}

\begin{document}
\title{ Flavon Magneto-Baryogenesis} 
\author{Fatemeh Elahi}
\email{felahi@ipm.ir}
\affiliation{School of Particles and Accelerators, Institute for Research in Fundamental Sciences IPM, Tehran, Iran}
\affiliation{PRISMA$^+$ Cluster of Excellence \& Mainz Institute for Theoretical Physics, \\
Johannes Glutenberg-Universität Mainz, 55099  Mainz, Germany} 
\author{Shiva Rostam Zadeh}
\email{sh\_rostamzadeh@ipm.ir}
\affiliation{School of Particles and Accelerators, Institute for Research in Fundamental Sciences IPM, Tehran, Iran}

\vspace*{0.5cm}
\begin{abstract}
\vspace*{0.5cm}

{
In this paper, we explore the evolution of baryon asymmetry as well as the hypermagnetic field in the early universe with an assumption that the flavon of the Froggatt-Nielsen carries an asymmetry. Through the decay of the flavon to Standard Model fermions, this asymmetry is transferred to fermions, where the right-handed electron keeps its asymmetry while its Yukawa interaction is out of thermal equilibrium. Through the existence of the flavon, we can ensure that the freezing-in temperature of the right-handed electron is closer to the electroweak phase transition than the Standard cosmology scenario. With this trick, the asymmetry in the right-handed electron is saved for a longer time. Moreover, the injection of the asymmetry to the right-handed electron is gradual, which helps the preservation of the asymmetry in the right-handed sector significantly. Due to the intimate relationship between fermion number violation and the helicity of the hypermagnetic field, some of the asymmetry is used to amplify the hypermagnetic field which itself helps to preserve the remnant asymmetry through keeping the Yukawa processes out of thermal equilibrium. We find the sweet region of the parameter space that can produce the right asymmetry in the baryons while generating a large hypermagnetic field by the time of the electroweak phase transition. 
} 
\end{abstract}
\maketitle
%*********************INTRODUCTION*********************** 
\section{ Introduction}
\label{sec:Intro}
%*********************INTRODUCTION*********************** 
One of the most intriguing questions of particle physics is the observation of matter-antimatter asymmetry. The observed asymmetry of the baryons is 
\beq
\eta_B \equiv \frac{n_B - n_{\bar B}}{s} \simeq 8.5 \times 10^{-11},
\eeq
with $s = 2\pi^2 g_\star T^3/45$  being the entropy density. The value of $\eta_B$ has been obtained by two orthogonal methods, one from the Big Bang Nucleosynthesis measurements~\cite{Cooke:2013cba} and another one from the Planck data~\cite{Ade:2015xua}, and they match miraculously. If the universe had started with an equal number of baryons as anti-baryons, three necessary and sufficient conditions known as Sakharov conditions are needed to generate a baryonic asymmetry: 1) Baryon number violation, 2) C and CP violation, and 3) out of thermal equilibrium process ~\cite{Sakharov:1967dj}. To explain the observed baryon asymmetry of the universe, physics beyond the Standard Model (SM) and new degrees of freedom are needed (e.g,~\cite{FUKUGITA198645,AFFLECK1985361}). Furthermore, it has been shown that baryon number violation is highly influenced by the presence of a hypermagnetic field~\cite{Semikoz:2011, Dvornikov:2013, Kuzmin:1985mm, Zadeh:2018usa,Long:2014,Rubakov:1986am,Giovannini:1998,Giovannini:1998b,Joyce:1997, Khlebnikov:1988sr,Zadeh:2016nfk,Zadeh:2015oqf,Mottola:1990bz}. That is because in the Standard Model (SM), baryon number violation is proportional to $E_Y \cdot B_Y$, where $E_Y$ and $B_Y$ are the hypercharge electric and magnetic fields, respectively.

Interestingly, there are some questions in the observations of widespread large scale magnetic fields in the Universe as well. Large scale magnetic fields in causally disconnect patches have been observed to have similar amplitudes~\cite{Kronberg:1993vk,Kulsrud:2008,Harrison:1973zz}. Even though part of the community believes that the origin of these magnetic fields is some astrophysical activities due to late post-recombination physics, some cosmologists insist that these observations roots in the early Universe. There exist different scenarios which try to explain the origin and the evolution of these cosmic magnetic fields which are referred to as magnetogenesis scenarios~\cite{Semikoz:2011, Dvornikov:2013,Quashnock:1988vs, Kibble:1995,Sigl:1997,Vachaspati:1991nm,Enqvist:1993,Enqvist:1994,Olesen:1997,Baym:1996,Grasso:2001,Neronov:2010,Neronov:2009,Tavecchio:2011,Tavecchio:2010,Giovannini:1998b,Joyce:1997,Wolfe:2008}. The evolution of the magnetic fields is not rigorously understood; however simple conservative estimates indicate that to justify the current magnetic fields, we need to have magnetic fields with amplitudes about $10^{20}$G by the electroweak phase transition (EWPT)~\cite{Dvornikov:2013,Fujita:2016,Giovannini:1997gp,Giovannini:1998,Giovannini:2013,Long:2016,Joyce:1997}. It is worth emphasizing that the quoted value is a rough estimate since there are numerous non-linear effects before and after the EWPT that have not been considered in this estimation.

In this paper, we are interested in scenarios where the initial seed of the hypermagnetic field amplitude (HMFA) is small, and through the existence of baryonic asymmetry, we get a large value of HMFA ($\sim 10^{20}$G) by the time of EWPT.  In Ref.~\cite{Zadeh:2018usa}, however, the authors have shown that in the standard cosmology (SC), this is rather impossible, even if we start with a large baryonic asymmetry. Therefore, we need to consider alternatives in the non-standard cosmology. 

To succeed in our mission, on the one hand, we need a mechanism that generates a large baryonic asymmetry that can be used to amplify a small seed of hypermagnetic field; on the other hand, we need to control the effect of the sphalerons-- either by changing the Hubble rate such that the freeze-in\footnote{The temperature at which the right-handed electron comes into thermal equilibrium.} of the right-handed electron occurs closer to the EWPT and/or by injection the asymmetry into right-handed electron slowly.

The importance of the right-handed electron is because of the following: If we insist on having the constraint $B-L=0$ and we have some initial asymmetry in the right-handed electron, then we must have some asymmetry in the baryonic sector as well. Right-handed electrons at high temperatures are not in thermal equilibrium and therefore cannot lose their asymmetry. However, once their Yukawa interaction's rate gets higher than the Hubble rate, then the asymmetry in the right-handed electron can be transferred into a left-handed electron and electron neutrino, and then weak sphalerons can wash out the asymmetry --eating the asymmetry preserving $B-L$ until it becomes zero. Before the EWPT, the rate of weak sphalerons is proportional to $T^4$, but then after the EWPT, their rate becomes increasingly more suppressed. Therefore, the rate of change in baryon asymmetry is more efficient before the EWPT. In the standard cosmology, the difference between the freeze-in temperature of right-handed electron, $T_R \simeq 10^5 \gev$, and the temperature at which EWPT occurs ($\TEW$) is large enough that the weak sphalerons have enough time to wash out the asymmetry. To avoid this problem, one solution is to change the cosmological evolution.

Recently, Chen et al.~\cite{Chen:2019wnk} discussed the generation of baryon asymmetry through the decay of the flavon. In this paper, the flavon dominates the energy density of the universe and causes the freeze-in of the right-handed electron to delay. Their scenario is motivated because the flavon of the Froggatt Nielsen (FN) is theoretically motivated to justify the hierarchy of fermion masses~\cite{Froggatt:1978nt,Weinberg:1979sa,Alvarado:2017bax}. The paper~\cite{Chen:2019wnk} has an obvious merit in that it explains two problems with a single theory. In this paper, we would like to be even more ambitious and find the region of the parameter space that can solve the magnetogensis as well. We find that only a small region of the parameter space can give satisfactory results, and that is with the assumption that the flavon only couples to the first generation of fermions. This way, the branching ratio of the flavon to the electron is more significant and thus more asymmetry can be transferred into the fermionic sector. Since the masses of the first generation are the most troublesome compared with the electroweak scale, we insist that this assumption is justifiable. 

Our results give the most desirable outcome when the cut off of the theory is about $10^{7.5} \gev$, and the mass of the flavon is nearly $15 \TeV$. The initial comoving wavenumber of the hyermagnetic field should also be about $0.5 \times 10^{-7}\times \T$, where $\T$ is when the flavon starts dominating.

The organization of the paper is as follows: In Section~\ref{sec:model}, we explain the FN mechanism and the couplings of the flavon with fermions. The nature of the FN symmetry as well as the evolution of the flavon in the early universe are discussed in Sections~\ref{sec:symmetry} and~\ref{sec:cosmo}, respectively. Section~\ref{sec:AMHD} is devoted to the evolution of the hypermagnetic field and Section~\ref{sec:asymmetry} discusses the Boltzmann equation of the right-handed electron in the presence of a flavon, sphaleron, and a non-zero small seed of the hypermagnetic field. In Section~\ref{sec:num}, we do a numerical study of the coupled Boltzmann equations. First, we discuss one benchmark in great detail, and then we scan through the parameter space and find the desired region. The concluding remarks are presented in Section~\ref{sec:conclusion}.

 %*********************Model *********************** 
\section{ Flavon Model }
\label{sec:model}
%*********************Model *********************** 
The Froggatt-Nielsen (FN) mechanism is a proposal to reproduce the mass hierarchy among the Standard Model (SM) fermions with $O(1)$ Yukawa couplings. The solution it proposes is charging the fermions under a new symmetry such that the lighter fermions have a larger charge. The charges of the fermions causes their Yukawa interactions to be modified. That is their Yukawa interactions at low energies become
\beq
\Lc_{\text{Yuk}} \supset y^f_{ij} \left(\frac{S_0}{\Lambda}\right)^{n_{ij}} \bar f_{L_i} \phi f_{R_j},
\label{eq:lag}
\eeq
where $\phi$ represents the Higgs, and $f_{L,R}$ are the SM left-handed and right-handed fermions, respectively. The indices $i,j$ represent the fermion's generations, and $n_{ij}$ is related to the FN charges of fermions. The complex scalar $S_0$, known as flavon, has a charge of $-1$ under the FN symmetry, and it is used to cancel the charges of the fermions in the Yukawa interactions.
In this set-up, Higgs does not have any FN charges. The cut-off scale $ \Lambda$ represents the mass of some vector-like fermions  at UV scales. Once $S_0$ acquires a vacuum expectation value (vev), the FN symmetry spontaneously breaks. After the FN spontaneous symmetry breaking (SSB), $S_0 $ obtains a dynamical part $S$ and a constant part $v_s$: $S_0 \to S + v_S$. The masses of fermions are the result of both Electroweak SSB \textit{and} FN SSB\footnote{We assume the Electroweak SSB to occur around $\TEW \simeq 160 \ \gev$. The FN SSB is expected to be at much higher temperatures, but its value is a free parameter that can be tuned.}, and are proportional to $ y_{ij} \left(\frac {v_S}{ \Lambda}\right)^{n_{ij}} v_\phi$, i.e. the SM Yukawa couplings of fermions are $ y_{ij} \left(\frac {v_S}{ \Lambda}\right)^{n_{ij}} $. Knowing the fermion masses and their FN charges,  $\epsilon \equiv v_S/\Lambda$ can be estimated, and in the most minimalistic scenario, it is approximately $0.2$. The purpose\footnote{It is important to mention that fermions, unlike Higgs, do not suffer from untamed quantum corrections. That is the radiative correction to their mass is always proportional to $v_h$ and thus it is finite.}  of the FN mechanism is to make the $y_{ij}$ in Eq.~\ref{eq:lag} natural, O(1).

We consider a scenario where the FN SSB occurs much earlier than the EWPT, and thus it is important to comment on the coupling of the dynamical field $S$ with $\bar f_{L_i} f_{R_j}  \phi$. We use the notation where this coupling is $g_{ij}/\Lambda$, with
\beq 
g_{ij}\equiv  y^f_{ij}  n_{ij} \epsilon^{n_{ij}-1} . 
\eeq
 
Furthermore, we focus on the case where only the first generation is charged under the FN symmetry.\footnote{Note that we do not have any off-diagonal entry in the couplings of the flavon with the SM fermions. In other words, the couplings of fermions with the flavon in the interaction basis are the same as the mass basis.} This is justified because the first-generation has the smallest masses in the SM. Specifically,  we will take the charges of the first generation as the following~\cite{Bauer:2016rxs}
\beq
Q_{\FN}(\bar Q_1, u_1, d_1,  \bar L_1, e_1) =  (3, 5,4, 5,4), 
\eeq
which using the definition $n_F = n_{\bar F_L} + n_{F_R}$ leads to 
\beq
n_e = 9 \hspace{0.3 in}  n_u = 8 \hspace{0.3 in} n_d = 7. 
\eeq

 %*********************Model *********************** 
\subsection{ The Nature of the FN Symmetry and the Generation of Flavon Asymmetry }
\label{sec:symmetry}
%*********************Model *********************** 
Thus far, we have not commented on the nature of the FN symmetry. In the following, we will discuss what kind of symmetries are suitable.  In general, the FN symmetry can be global/local and continuous/discrete. Given that the FN symmetry is severely anomalous, we focus on the global case. As a result of SSB of continuous global symmetry, a massless Goldstone boson emerges; a consequence that is strongly disfavored by CMB~\cite{Banerjee:2016suz,Eisenstein:1999,cuesta2015neutrino}. To avoid this problem, we can assume the FN symmetry is discrete, $\Z_N$~\cite{Lillard:2018}, where we take $N= 20$ to make sure the charges of light fermions are well-defined. Even though the SSB of a discrete symmetry leads to the production of domain walls in the early Universe, the lack of observation of domain walls so far can be cosmologically justified (see Ref.~\cite{Witten:1997ep,Preskill:1991kd,Abel:1995wk,Lazarides:1982tw} for more information).  After the FN SSB, both the real and the imaginary components of $S$ gain different non-zero masses. However, as argued in Ref.~\cite{Chen:2019wnk}, one could start with more complex fields and the flavon can be defined as a complex linear combination of these fields with the same mass. The sameness of the mass of these degrees of freedom can be protected by a symmetry such as a custodial symmetry~\cite{Hambye:2009,Arcadi:2016}. Defining $S$ as a complex linear combination of the scalars means that $S$ can carry some initial asymmetry (e.g., through Affleck-Dine mechanism~\cite{Affleck:1984fy, Chen:2019wnk}). 

It should be noted that the non-renormalizable interactions of the flavon, e.g.,
\beq 
V(S) =  \frac{\kappa}{\Lambda^{N-4}} S^N + \frac{\kappa'}{\Lambda^{N-4}} S^{*N}+ V'(S S^*), 
\eeq
with $\kappa^* \neq \kappa'$, are responsible for the generation of flavon asymmetry~\cite{Kitano:2008tk}. Thereby, their effect is more relevant at high temperatures, when the suppression of $T/\Lambda$ is smaller; but they are irrelevant at lower temperatures. Here, we assume a positive asymmetry in the flavon is generated at high temperatures.

 %*********************Model *********************** 
\subsection{The Cosmology of the Flavon}
\label{sec:cosmo}
%*********************Model ***********************

In our scenario, we need the flavon to have a large asymmetry at high temperatures. This asymmetry should be conserved until the flavon starts its coherent oscillation. During this epoch, the flavon decays to fermions through $S \to f_{L_1}  f_{R_1} \phi$, and its asymmetry penetrates to the fermionic sector. In the following, we will discuss each of these steps in greater detail.

A weakly interacting scalar field goes through coherent oscillation for a period of $ H \lesssim m_S$.  That is for temperatures below $ \sqrt{ M_H m_S}$, where $M_H = \Mpl/(1.66 \sqrt{g_*}) \simeq 1.4 \times 10^{17} \gev$ is the reduced Planck mass and $m_S$ is the flavon mass. It is worth saying that lighter flavons have larger amplitudes of oscillation and thus enjoys higher yield~\cite{Lillard:2018}.  

In order to have a successful coherent oscillation, we must make sure that the production of the flavon is out of equilibrium~\cite{Lillard:2018}. Thereby, we require
\beq
n\langle \sigma v \rangle_{f_{L_1}  f_{R_1} \to  S H} \simeq  \zeta(3) \frac{T^3}{4\pi^3} \frac{\sum_{f = e,u,d}|g_f|^2}{\Lambda^2}   <H.   
\eeq
This condition ensures that excited states of flavons, which would have messed up their coherency, do not get produced. 
Let us define the temperature at which the rate of the flavon production equals to Hubble rate as $T_M$: 
\beq 
T_M (\Lambda) \equiv \frac{4 \pi^ 3 \Lambda^2}{\zeta(3) M_H \sum_{f = e,u,d}|g_f|^2}. 
\label{eq:Tmax}
\eeq
In order to obtain the above equation, we have used the Effective Field Theory (EFT) approach, and thus the maximum temperature of the Universe ($\Tmax$) must be smaller than $\Lambda$. Therefore, we require $\Tmax(\Lambda) \equiv \text{min}\, [T_M (\Lambda), \Lambda]$.

During the coherent oscillation, the flavon redshifts like cold matter (i.e.\ $\rho_S \propto a^3$). Consequently, at some temperature $\T$, the energy density of the flavon equals that of the radiation: $\left. \rho_S\right|_{\T} = \left. \rho_{\text{rad}}\right|_{\T}$. For $T < \T$, $\rho_S$ dominates the energy density of the Universe, and thus the Hubble rate gets modified. During this epoch, the flavon decays to $ff\phi$ which contributes to the radiation of the universe, increasing $\rho_{\text{rad}}$, and eventually leading to the termination of the matter-domination. The evolution equations of $\rho_S$ and $\rho_{\text{rad}}$ are as follows~\cite{Chen:2019wnk}:
\begin{align}
&\dot \rho_S + 3H \rho_S = -\Gamma_S \rho_S \nonumber\\
&\dot \rho_{\text{rad}} + 4H \rho_{\text{rad}} = \Gamma_S \rho_{S},
\label{eq:density}
\end{align}
with 
\beq
\Gamma_S = \frac{\sum_{f = e,u,d}|g_f|^2 }{64 \pi^3 \epsilon^2}\frac{m_S^3}{\Lambda^2}
\eeq 
being the total decay of the flavon ($S \to f_{L_1} f_{R_1} H$), and $$H =\sqrt{ \frac{8\pi}{3 \Mpl^2} (\rho_S + \rho_{\text{rad}})},$$ being the Hubble rate. 

\begin{figure}
\centering 
\includegraphics[width=0.52\textwidth, height=0.42\textheight]{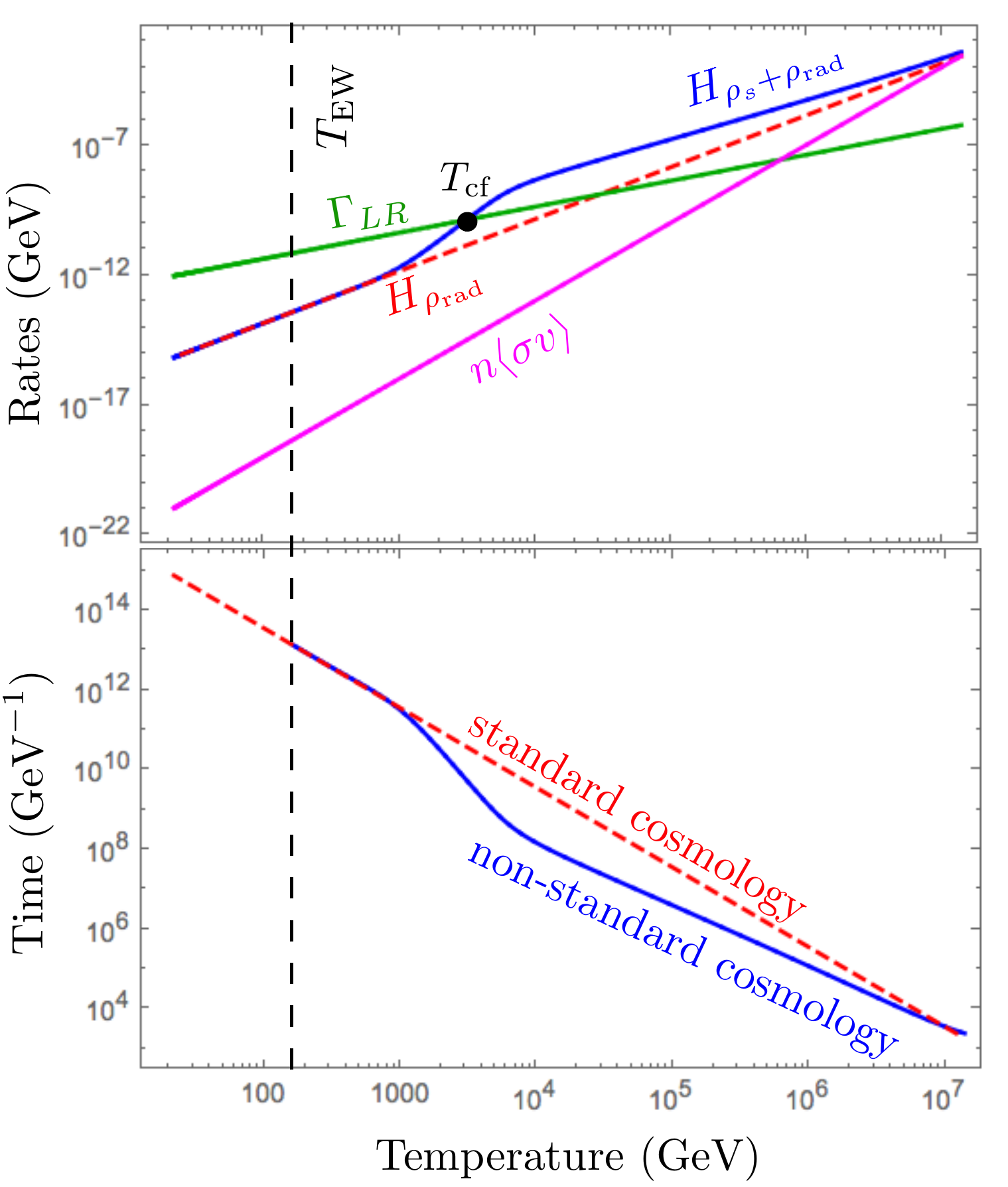}
\caption{The Hubble rates both in our scenario $ H_{\rho_S + \rho_{\text{rad}}}$ (shown in solid blue) and in the standard cosmology $ H_{\rho_{\text{rad}}}$ (dashed red), as well as the rate of electron Yukawa interaction $\Gamma_{LR}$ (solid green) are shown in the upper panel. As the plot demonstrates, $\Tx$ is much closer to the EW temperature in the presence of a flavon than the standard cosmology.  For the demonstration, we have also shown the rate of production of flavon $n \langle \sigma v \rangle$ (in magenta) to ensure that the flavon is indeed out of equilibrium and thus goes through coherent oscillation and redshifts like matter. The relation between time and temperature in the non-standard (solid blue) and standard (dashed red) cosmology is shown in the lower panel and it demonstrates once the flavon decays, we are back to standard cosmology.  This plot is with the assumption that $ \T = 1.4 \times 10^6 \gev$. }
\label{fig:rates}
\end{figure}

The analytical approximate solutions of Eqs.~\ref{eq:density} for the time interval $ t_\star <t < \Gamma_S^{-1}$ are the following~\cite{Gorbunov:2011zz} :
\begin{align}
&\rho^{a}_S(t) \simeq \frac{\Mpl^2}{6 \pi t^2} e^{-\Gamma_S t}\nonumber\\
&\rho^a_{\text{rad}} (t) \simeq \frac{ \Mpl^2 t_\star^{2/3}}{6 \pi t^{8/3}} + \frac{ \Gamma_S \Mpl^2}{ 10 \pi t}, 
\label{eq:rhoin}
\end{align}
where $t_\star$ is the time that corresponds to $\T$. To convert between temperature and time, we use the definition of the temperature, which is 
\beq
T = \left( \frac{\rho_{\text{rad}} (t)}{\frac{\pi^2}{30} g_{\star}}\right)^{1/4}.
\eeq
In order to obtain $t_\star$ for a given $\T$, we simply plug in the analytical solution, $\rho^a_{\text{rad}} (t_\star)$ into the above equation, and solve for $t_\star$. Eqs.~\ref{eq:rhoin} are also used to get $\rho_{S (\text{rad})} (t_\star) = \rho^a_{S (\text{rad})} (t_\star) $, which are needed as the initial conditions for solving Eqs.~\ref{eq:density} numerically. Other than the aforementioned tasks, we do not rely on the analytical solutions (Eqs.~\ref{eq:rhoin}) anymore. The numerical solution of $\rho_S$ and $\rho_{\text{rad}}$ as a function of temperature, assuming $\T = T_{max}(\Lambda =10^8 \gev)$, are shown in Fig.~\ref{fig:rates} -- upper panel. The lower panel compares the time-temperature conversion in the SC and our scenario which includes an intermediate matter domination (non-standard cosmology).

As it is apparent in Eq.\ \ref{eq:lag}, the flavon interactions with SM particles respect $B$ and $L$ symmetries, and therefore the $B-L$ symmetry, that is respected in the framework of the SM as well. As a result of the flavon and antiflavon decaying to SM fermions and antifermions, the flavon-antiflavon asymmetry is transferred to left-right asymmetry in the SM content. The left-right asymmetry produced in the quark sector is washed out immediately by the strong sphalerons. However, in the leptonic sector, the produced asymmetry in the right-handed electron is preserved above a critical temperature.\footnote{Soon it will be clarified that this is the temperature at which the chirality flip rate of the electron becomes equal to the Hubble rate.} Therefore, the weak sphalerons, which are only active before the EWPT and act only on left-handed particles, partially convert the asymmetry of left-handed leptons into a baryon (B) asymmetry~\cite{Rubakov:1996}. Consequently, we gain a simultaneous asymmetry in the quark and lepton sectors. Indeed, weak sphalerons tend to wash out the asymmetry of these two sectors. However, the washout process is successful if and only if all of the Yukawa interactions are in thermal equilibrium~\cite{Campbell:1992,Cline:1993,Cline:1994,Harvey:1990qw,Gorbunov:2011zz,Rubakov:1996,Kuzmin:1985mm,bodeker2019equilibration}. The rate of the Yukawa interactions is proportional to $y_f^2 T$, where $y_f$ is their SM Yukawa coupling. Since electrons have a small Yukawa coupling, they are the last fermions\footnote{Here, we consider that the neutrinos are massless such as in the SM.} that enter thermal equilibrium, and thus the action of weak sphalerons is limited by electron's chirality flip process~\cite{Campbell:1992,Cline:1993,Cline:1994}. Specifically, it is the right-handed electron that plays a key role in preserving the asymmetries.

Due to the importance of the chirality flip of the right-handed electron, its rate has been extensively studied, and the most recent calculation of it is~\cite{bodeker2019equilibration,Kamada:2016}: 
\beq
\Gamma_{LR} \simeq 10^{-2} y_e^2 T.
\label{eq:GamLR}
\eeq

Let us define the temperature at which the chirality flip of the right-handed electron process goes to equilibrium as
$\Tx \equiv \left. T\right|_{\Gamma_{LR} \sim H}.$ In the SC, $\Tx$ is about $(10-100)  \TeV$, as can be seen in Fig.~\ref{fig:rates}.\footnote{The intersection of the dashed red line and solid green line in the upper panel.} Even though the asymmetries are preserved up to this temperature, it has been shown that below $\Tx $ the weak sphalerons still have enough time to wash out the asymmetries due to their high rates~\cite{Campbell:1992,Cline:1993,Cline:1994}. In this scenario, however, the presence of the flavon may change the story~\cite{Chen:2019wnk}, because 
\begin{itemize}
\item it brings $\Tx$ relatively closer to $\TEW$, and  
\item it transfers the asymmetry to the fermionic sector {\it gradually}.\footnote{The authors of  Ref.~\cite{Chen:2019wnk} considered the decay of the flavon to tau and electron, which leads to a much larger decay width of the flavon compared to that of our scenario. Since the gradual decay of the flavon is more important for our scenario, we considered the decay of the flavon only to the first generation of fermions.}

\end{itemize}
In this project, we are not only interested in acquiring the right baryonic asymmetry of the Universe, but also we want the asymmetries to amplify a small seed of the hypermagnetic field to amplitudes as large as $10^{20}$G at the onset of the EWPT\footnote{Here we neglect the possible change of the baryonic asymmetry during EWPT.}.  It has been argued that a hypermagnetic field with this amplitude at $\TEW$ can lead to the observed magnetic fields as large as $10^{-17}-10^{-15}$G observed in the intergalactic medium (IGM). Thereby, in this paper, we are interested in the region of the parameter space that yields 
\begin{align}
\eta_B(\TEW) &\simeq 8.5 \times 10^{-11} \nonumber\\
 B_Y(\TEW)& \gtrsim 10^{19}\textrm{G}.
 \label{eq:desired} 
\end{align}

Before the EWPT, the evolution of hypermagnetic fields and the asymmetries are strongly intertwined through Abelian anomaly ($\partial_\mu J^\mu_{B,L} \propto \vec  E_Y \cdot \vec B_Y$) and chiral magnetic effect (CME)~\cite{Laine:2005,Zadeh:2018usa,Zadeh:2016nfk,Zadeh:2015oqf,Appelquist:1981vg,Kajantie:1996}. These effects, together, ensure the conversion of the asymmetries to the helicity of hypermagnetic fields, and vice-versa. However, it has been shown that in the framework of the SM and the presence of the weak sphalerons, the initial asymmetries are rapidly washed out and no growth of the hypermagnetic field happens~\cite{Zadeh:2018usa}. Indeed, the growth can happen if the asymmetry is somehow preserved for a longer time compared to SC~\cite{Zadeh:2018usa}; a task that is achievable in our model through the flavon\footnote{The arising hypermagnetic field, can in return, push the Yukawa interactions out of equilibrium, assisting the preservation of the asymmetry. The deviation of the Yukawa interactions from equilibrium is highly correlated with their Yukawa rate: the slower the rate, the larger the deviation from equilibrium. Therefore, the effect of the hypermagnetic field is particularly important for the chirality flip of the electrons~\cite{Zadeh:2018usa}.
}. In the following section, we will look at the evolution equations of the hypermagnetic fields.  

 %*********************EVOLUTION*********************** 
\section{ Anomalous Magnetohydrodynamics }
\label{sec:AMHD}
%*********************EVOLUTION*********************** 
In the static limit, the effective action of the soft $U(1)_Y$ gauge fields can be derived via the method of dimensional reduction~\cite{Laine:2005,Appelquist:1981vg,Kajantie:1996}. The corresponding Lagrangian describing the dynamics of these fields at finite fermionic density in the Minkowski spacetime is the following~\cite{Laine:2005,Joyce:1997,Zadeh:2015oqf,Kajantie:1996}:
\beq 
\Lc = - \frac{1}{4} Y_{\mu \nu} Y^{\mu \nu} - J_Y^\mu Y_\mu - c'_E \frac{\alpha'}{8 \pi} ( 2 Y\cdot B_Y),
\label{eq:AMHDlag}
\eeq
where $\alpha' \equiv g^{'2}/4\pi \simeq 0.01$ is the fine structure constant of the hypercharge interaction. In Eq.~\ref{eq:AMHDlag}, the first term is the kinetic term of the hypercharge field, $J_Y$ is the Ohmic current, and the last contribution is related to the Chern-Simons term, which leads to the CME~\cite{Laine:2005}. The Chern-Simons coefficient, $c'_E$, can be written as~\cite{Zadeh:2015oqf,Laine:2005}
\begin{align}
c'_E = \sum_{i=1}^{n_G} &\left[-2 \mu_{R_i} + \mu_{L_i} - \frac{2}{3} \mu_{d_{R_i}}- \frac{8}{3} \mu_{u_{R_i}} + \frac{1}{3} \mu_{Q_i}\right],
\label{eq:ceprime}
\end{align}
where the $\mu$'s are the chemical potentials of various chiral fields, and $n_G$ is the number of generations. Let us make the simplifying assumption that all Yukawa interactions, other than that of the electron, are in thermal equilibrium.  Thereby, we can obtain all of the chemical potentials in terms of the chemical potential of the right-handed electron by requiring $B/3 - L_i$ (with $i$ being the generation index) conservation as well as the hypercharge neutrality in the plasma. As a result, $c'_E$ can be reduced to $c'_E = - 99/37 \mu_{e_{R}}$. Furthermore, one important chemical potential that has observational significance is  $\mu_B= \sum^{n_G}_{i=1} \left[2  \mu_{Q_i} +\mu_{u_{R_i}} + \mu_{d_{R_i}}\right]$. Using the aforementioned simplifying assumptions and conservation laws, we obtain $\mu_B = \frac{198}{481} \mu_{e_R}$~\cite{Zadeh:2018usa,Chen:2019wnk}. 

Since we are interested in studying the evolution equation of the hypermagnetic field in the early Universe, we must consider the Friedman-Robertson-Walker (FRW) metric. Therefore, the Lagrangian in Eq.~\ref{eq:AMHDlag} will be slightly modified (see Appendix A in Ref.~\cite{Abbaslu:2019yiy}), and the resulting AMHD equations in the curved spacetime become the following:
\begin{align}
&\frac{1}{a} \vec \nabla\cdot \vec E_Y = 0, \hspace{0.5 in} \frac{1}{a} \vec \nabla \cdot \vec B_Y = 0\hspace{0.3 in}
\label{eq:Maxwell} 
\end{align}
\begin{align}
 &\partial_t \vec B_Y+ 2H \vec B_Y = - \frac{1}{a} \vec \nabla \times \vec E_Y\hspace{0.7 in}
 \label{eq:BY}
 \end{align}
 \begin{align}
  &\vec J_{\text{Ohm}} = \sigma (\vec E_Y + \vec v \times \vec B_Y)\hspace{0.95 in}
  \label{eq:johm}
  \end{align}
  \begin{align}
 &\vec J _{\text{cm}} = - \frac{\alpha'}{2 \pi} c'_E \vec B_Y\hspace{1.35 in}
   \label{eq:jcm}
 \end{align}
 \begin{align}
 &\vec J_{\text{Ohm}} +  \vec J _{\text{cm}} = \frac{1}{a} \vec \nabla \times \vec B_Y - \left(\partial_t \vec E_Y + 2 H \vec E_Y\right),
  \label{eq:j} %new label
\end{align}
 where $\sigma \simeq 100 T$ is the electrical hypercoductivity of the plasma, $H= \dot a/a$ is the Hubble parameter, $a$ is the scale factor, and the currents $\vec J_{\text{Ohm}}$ and $\vec J_{\text{cm}}$ are the Ohmic and chiral magnetic currents, respectively. The latter current, which is in the direction of the hypermagnetic field, comes from the Chern-Simons term and promotes the ordinary magnetohydrodynamics equations to anomalous magnetohydrodynamics (AMHD) equations. The terms containing the Hubble parameter $H$ are related to the expansion of the Universe. Using Eqs.~\ref{eq:j} and~\ref{eq:johm} and neglecting the displacement current ($\partial_t  \vec E_Y + 2 H \vec E_Y$) in the lab frame, the hyperelectric field will be obtained as 
 \beq
 \vec E_Y = \frac{1}{a \sigma} \vec \nabla \times \vec B_Y + \frac{\alpha'}{2 \pi \sigma} c'_E \vec B_Y - \vec v \times \vec B_Y. 
 \label{eq:EY1}
\eeq
In the above equation, we can neglect the last term containing the velocity of the plasma. That is because the correlation distance of the hypermagnetic field is much larger than the length scale of the variation of the bulk velocity. Therefore, the hypercharge infrared modes are practically unaffected by the plasma velocity~\cite{Rubakov:1986am}.

Replacing Eq.~\ref{eq:EY1} in Eq.~\ref{eq:BY}, we can solve for the evolution equation of the hypermagnetic field:
 \beq
 \partial_t \vec B_Y + 2 H \vec B_Y = \frac{1}{a^2 \sigma} \nabla^2 \vec B_Y - \frac{\alpha'}{2 \pi a \sigma} c'_E  \vec \nabla \times \vec B_Y.
 \eeq

 Since $\vec \nabla \cdot \vec B_Y =0$, we can write the hypermagnetic field as $\vec B_Y = (1/a) \nabla \times \vec A_Y$, where $\vec A_Y$ is the vector potential. Considering a fully helical hypermagnetic field, the following non-trivial Chern-Simons wave configuration for $\vec A_Y$ can be chosen~\cite{Dvornikov:2013, Zadeh:2018usa,Giovannini:1998,Giovannini:1998b,Zadeh:2016nfk,Zadeh:2015oqf}:
 \beq 
 \vec A_Y = \gamma (t) (\sin kz, \cos kz, 0),
 \eeq
 where $\gamma(t)$ is the time-dependent amplitude of $\vec A_Y$, and $k$ is the comoving wave number. Using this configuration, the hypermagnetic field becomes $\vec B_Y = (1/a) k \vec A_Y$, and consequently 
 \begin{align}
& \vec E_Y =\frac{k'}{\sigma} \vec B_Y + \frac{\alpha'}{2 \pi  \sigma} c'_E  \vec B_Y,
\label{eq:EY}
\end{align}
and
\begin{align}
& \partial_t \vec B_Y + 2 H \vec B_Y = -\frac{k^{'2}}{\sigma} \vec B_Y - \frac{\alpha'}{2 \pi \sigma} c'_E  k' \vec B_Y,
\label{eq:BYdiff}
 \end{align}
 with $k' \equiv k/a = k T$, can be derived. Let us define the amplitude of the hypermagnetic field ($\vec B_Y$) as  $B_Y (t) \equiv k' \gamma (t)$. Hence, Eq.~\ref{eq:BYdiff} can be rewritten as the following
 \beq
\partial_t  B_Y + 2 H B_Y = -\frac{k'}{\sigma}  B_Y (k' + \frac{\alpha'}{2 \pi } c'_E  ).
\label{eq:By}
 \eeq

Thus far, we have seen that if $\mu_i \neq 0$ (there is a non-zero asymmetry), $E_Y(t)$ and $B_Y(t)$ get modified due to the Chern-Simons term. The evolution of asymmetries, on the other hand, depends on $E_Y \cdot B_Y$. Therefore, the modified electric field and hypermagnetic field become important in the evolution of asymmetries. In the following section, we will discuss how this effect shows up in the evolution of asymmetries in greater detail. 

%*********************EVOLUTION***********************
\section{Evolution of matter asymmetries }
\label{sec:asymmetry}
%*********************EVOLUTION*********************** 
As mentioned earlier, with the simplifying assumptions that we have made, all matter asymmetries can be obtained in terms of the asymmetry of the right-handed electron. Therefore, it suffices to study the dynamics of this asymmetry, only~\cite{Zadeh:2018usa}. The asymmetry in the number density of the right-handed electrons 
%($n_R = n_{e_R} - n_{\bar e_R}$) 
can be found by solving the following Boltzman equation:
\begin{align}
\dot n_{e_R} &+  3 H n_{e_R}  =  - \Gamma_{LR}  \left(n_{e_R} - n_{e_L} + \frac{n_\phi}{2}\right) \nonumber\\
&+ B_e\Gamma_S n_S+ \frac{ \alpha' }{ \pi} \vec E_Y \cdot \vec B_Y. 
\label{eq:erasym}
\end{align}
In the above equation, $n_i$ with $ i=\{ e_R, e_L, \phi, S\}$, is the \textit{difference} between the number densities of a particle and its antiparticle. The term involving $H$ is due to the expansion of the Universe, and the term containing $\Gamma_{LR}$ shows the effect of the electron Yukawa interaction. Note that the factor of $1/2$ in the parentheses is due to the spin statistics of the Higgs. Furthermore, the term $B_e\Gamma_S n_S$ comes from the decay of the flavon, with $B_e$ being the flavon branching ratio to electrons:
\beq
B_e = \frac{ g_e^2}{\sum_{f} g_f^2}, \hspace{0.2 in} f = e,u,d.
\eeq
Instead of $n_S$, it is more convenient to work with $\rho_S$. Therefore, we define a dimensionless parameter $\xi_S$ as $\xi_S \equiv n_S m_S/\rho_S$, which does not depend on time.
It should be noted that $\xi_S$ is different from the canonical definition of $\eta_S \equiv n_S/s$, where $s$ is the entropy density.

One important difference between our work and Ref.~\cite{Chen:2019wnk} is due to the term containing $\vec E_Y \cdot \vec B_Y$ in Eq.~\ref{eq:erasym}. This term comes from the Abelian anomaly equation:
\beq
\partial_\mu J^\mu_{e_R} =  - \frac{1}{4} Y_R^2 \frac{\alpha'}{4 \pi} Y_{\mu \nu} Y^{\mu \nu} = \frac{\alpha'}{\pi} \vec E_Y \cdot \vec B_Y,
\eeq
where $Y_R = -2$ is the hypercharge of the right-handed electron. The above equation relates the evolution of number densities to that of the helicity of the hypermagnetic field. Using Eq.~\ref{eq:EY}, we can derive 
 \beq
 \vec E_Y \cdot \vec B_Y  = \frac{ B^2_Y}{\sigma} \left( k' + \frac{\alpha'}{2 \pi }c'_E \right).
\label{eq:EdotB}
 \eeq
 As can be seen, the CME is not only important for the evolution of the hypermagnetic field as discussed in the previous section, but also it has a non-trivial effect on the evolution of the asymmetries via the term containing $c'_E$. Previously, we had defined $c'_E$ in terms of the chemical potential of right-handed electron: $c'_E = - 99/37 \mu_{e_{R}}$. We can convert $\mu_{e_R}$ to $n_{e_R}$ using $n_{e_R} = \mu_{e_R} T^2/6$.
 
In the subsequent section, we solve the coupled differential equations for $ \rho_S, \rho_{\text{rad}}$ (Eq.~\ref{eq:density}), $B_Y $ (Eq.~\ref{eq:By}), and $n_{e_R}$ (Eq.~\ref{eq:erasym}) numerically. To fully comprehend different stages of the evolutions, we first discuss one specific benchmark. We then move on to scanning the parameter space to find the desired region of the parameter space.

  %*********************EVOLUTION***********************
\section{ Numerical study }
\label{sec:num}
%*********************EVOLUTION*********************** 

In this section, we do a numerical study of the coupled evolution equations of $ \rho_S, \rho_{\text{rad}}, B_Y, $ and $n_{e_R}$ from $\T$ up to $\TEW$. Our free parameters are $ \T,  ~ m_S, ~\Lambda,~ \xi_S, ~ k, ~ B_Y(\T)$. Before diving into the numerical analysis, let us make a few comments on these parameters:
\begin{itemize}
  \item[$-$] We need the flavon production to stay out of equilibrium during coherent oscillation. Hence, the maximum value of $\T$ should be $\Tmax(\Lambda)$, as defined earlier. 
 \item[$-$] By looking at the evolution equations, we see that the ratio of $m_S/\Lambda$ is a recurring variable. Thereby, we find it more convenient to work with $\epsilon_m \equiv m_S/\Lambda$, and $\Lambda$ instead of $m_S$ and $\Lambda$. In order to respect EFT, we require $\epsilon_m \ll 1$. 
 \item[$-$] It has been shown that for $k \gtrsim  10^{-7}$, the hypermagnetic field does not survive the Ohmic dissipation in the plasma~\cite{Dvornikov:2013}. In our numerical analysis, we re-scale $k$ and work with $ c_{k_0} \equiv k/10^{-7}$, instead. 
 \item[$-$] As can be seen in Eq.~\ref{eq:By}, a non-zero initial seed is needed for the hypermagnetic field to be later amplified as a result of the CME.\footnote{The creation of the seed is beyond the scope of this study. Interested readers are encouraged to look into Ref.~\cite{Miranda:1998ne,Tsagas:2001ak,Enqvist:2004yy,Ashoorioon:2004rs,Hanayama:2005hd,Kunze:2005ef,Semikoz:2007ti,Abbaslu:2019yiy,Hanayama:2009as,Subramanian:2019jyd} on some of the possible mechanisms for the production of this seed.}  Here, we fix the initial amplitude of the hypermagnetic field to a small value of $B_Y (\T) = 0.01$G. 
 \item[$-$] In our scenario, we need large matter asymmetries in order to obtain the desired value of $B_Y(\TEW) \gtrsim 10^{19}$G , as explained earlier. Since the flavon is responsible for the generation of these asymmetries, we fix $\xi_S$ to its maximum value: $\xi_S =1$.
 \item[$-$] We further assume that all initial asymmetries in the Fermionic sector are zero (i.e, $\eta_f (\T) = 0$). 
\end{itemize} 

According to the above assumptions, the free parameters we work with in this paper, are 
$$ \T, \hspace{0.2 in} \epsilon_m, \hspace{0.2 in} \Lambda, \hspace{0.2 in} c_{k_0}. $$ 

  \begin{figure}[h!]
\includegraphics[width=0.5\textwidth, height=0.42\textheight]{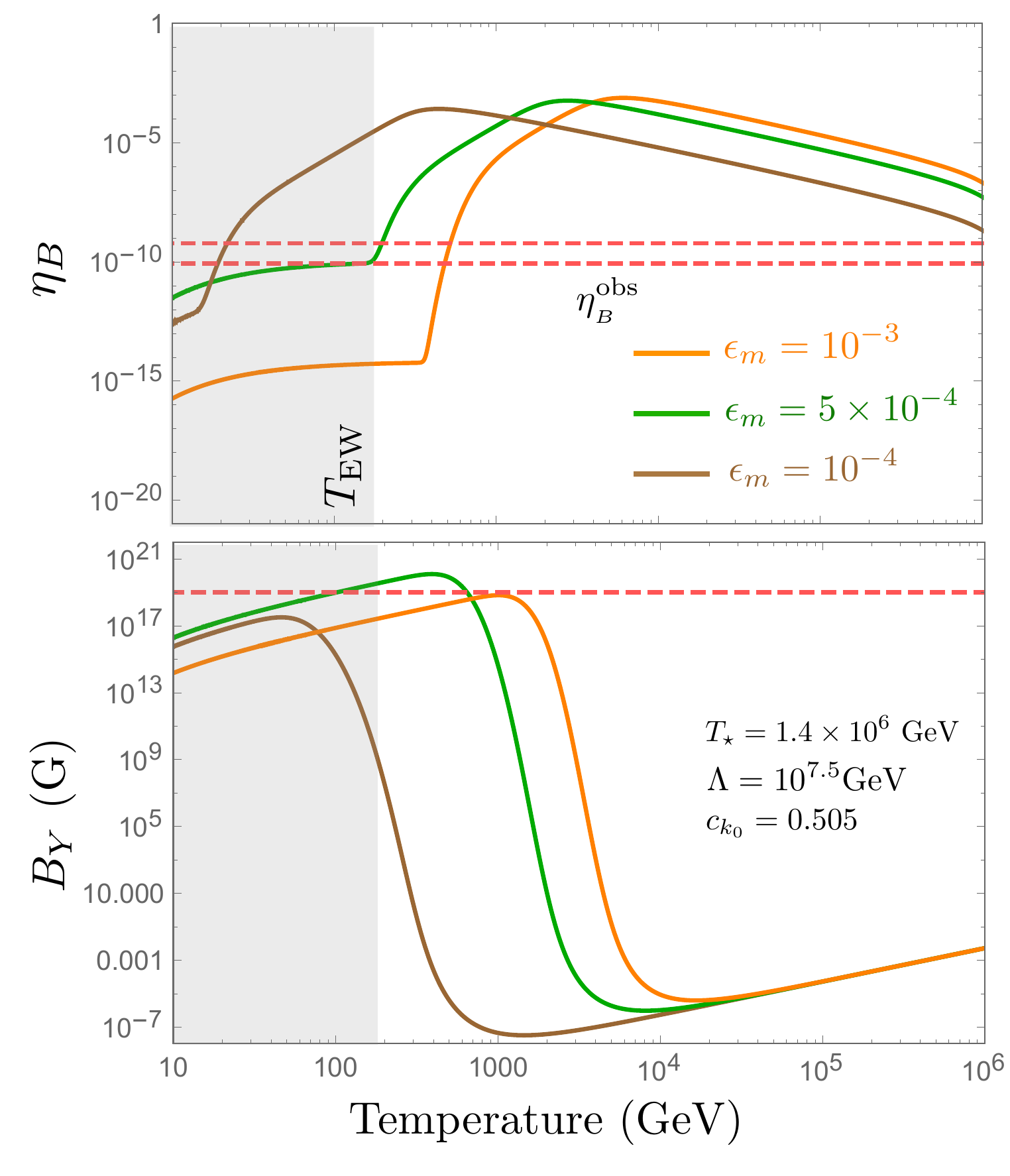}
\caption{The evolutions of $\eta_B $ (upper panel) and $B_Y$ (lower panel) as a function of temperature for $\epsilon_m = 10^{-3}$ (Orange), $\epsilon_m = 5 \times 10^{-4}$ (Green), and $\epsilon_m =10^{-4}$ (Brown) are presented. These plots are for the fixed values of $\Lambda = 10^{7.5} \gev$, $\T = 1.4 \times 10^{6} \gev$, and $c_{k_0} = 0.505$. As a result of the flavon decay at high temperatures, the asymmetries grow. Once $\eta_B$ reaches a large enough value, the hypermagnetic fields start getting amplified at the expense of eating some of the asymmetries. This process continues until the expansion rate of the universe becomes more important than the CME.} 
\label{fig:Tempevol}
\end{figure} 
%*********************EVOLUTION***********************
\subsection{ A Case Study}
\label{sec:case}
%*********************EVOLUTION*********************** 
In this subsection, we present a careful study of the evolution of $\eta_B(T)$ and $B_Y(T)$ as a function of temperature for the following benchmark:
\begin{align}
 &\Lambda = 10^{7.5} \gev, \epsilon_m =  5 \times 10^{-4}, \nonumber\\
& \T  \simeq 1.4 \times 10^{6} \gev, c_{k_0} = 0.505,
 \label{eq:benchmark}
 \end{align}
 where $\T$ is chosen as $T_\textrm{max}(\Lambda)$ for $\Lambda= 10^{7.5} \gev$.  The above parameters are chosen such that the desired values of $\eta_B(\TEW) \simeq 10^{-10}$ and $B_Y(\TEW) \gtrsim 10^{19}$G are obtained. To see how other benchmarks may change the results, we present the plots of $\eta_B(T)$ and $B_Y(T)$  for three different values of $\epsilon_m = 10^{-3},~5\times 10^{-4},~10^{-4}$ in Fig. \ref{fig:Tempevol};
the values of $\Lambda,~\T$, and $c_{k_0}$ are fixed as Eq.~\ref{eq:benchmark}.\footnote{It is worth mentioning that $\eta_B$ and $B_Y$ are highly sensitive to the exact value of $c_{k_0}$, and thus its value should be carefully tuned, as will be shown in the next subsection.} 

\begin{figure}[h!]
\includegraphics[width=0.52\textwidth, height=0.42\textheight]{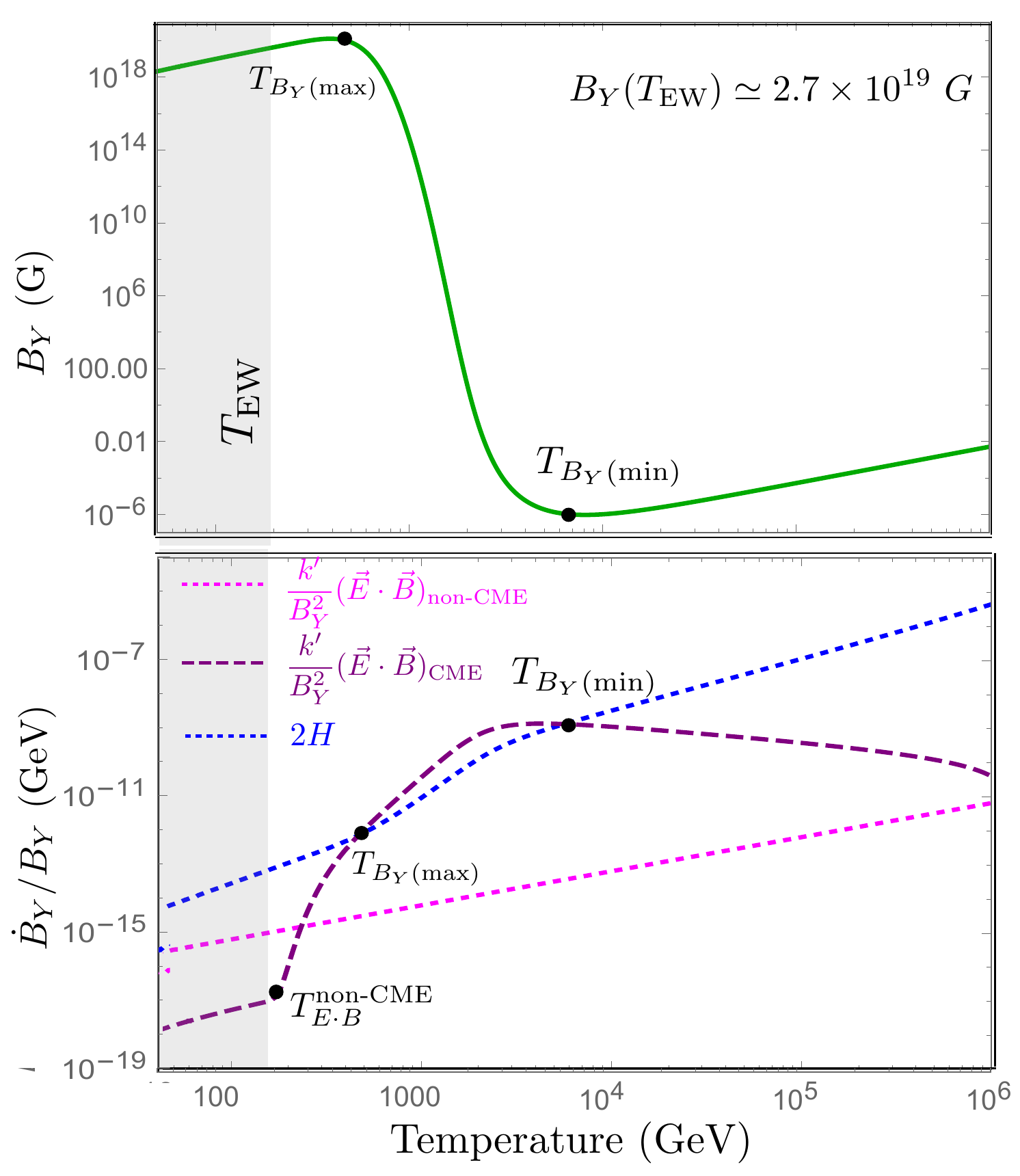}
\caption{
The evolution of $B_Y$ as a function of temperature is illustrated in the upper panel, while the lower panel shows the terms contributing to $\dot B_Y/B_Y$. The CME component (shown by dashed purple) leads to an increase in the HMFA, and the non-CME component (dotted magenta) reduces $B_Y$. We see that once the $(\vec E_Y \cdot \vec B_Y)$contribution exceeds the Hubble rate (dotted blue), the hypermagnetic field starts increasing and this trend continues until it falls below the Hubble rate again. } 
\label{fig:By2}
\end{figure}

\begin{figure}[h!]
\includegraphics[width=0.52\textwidth, height=0.44\textheight]{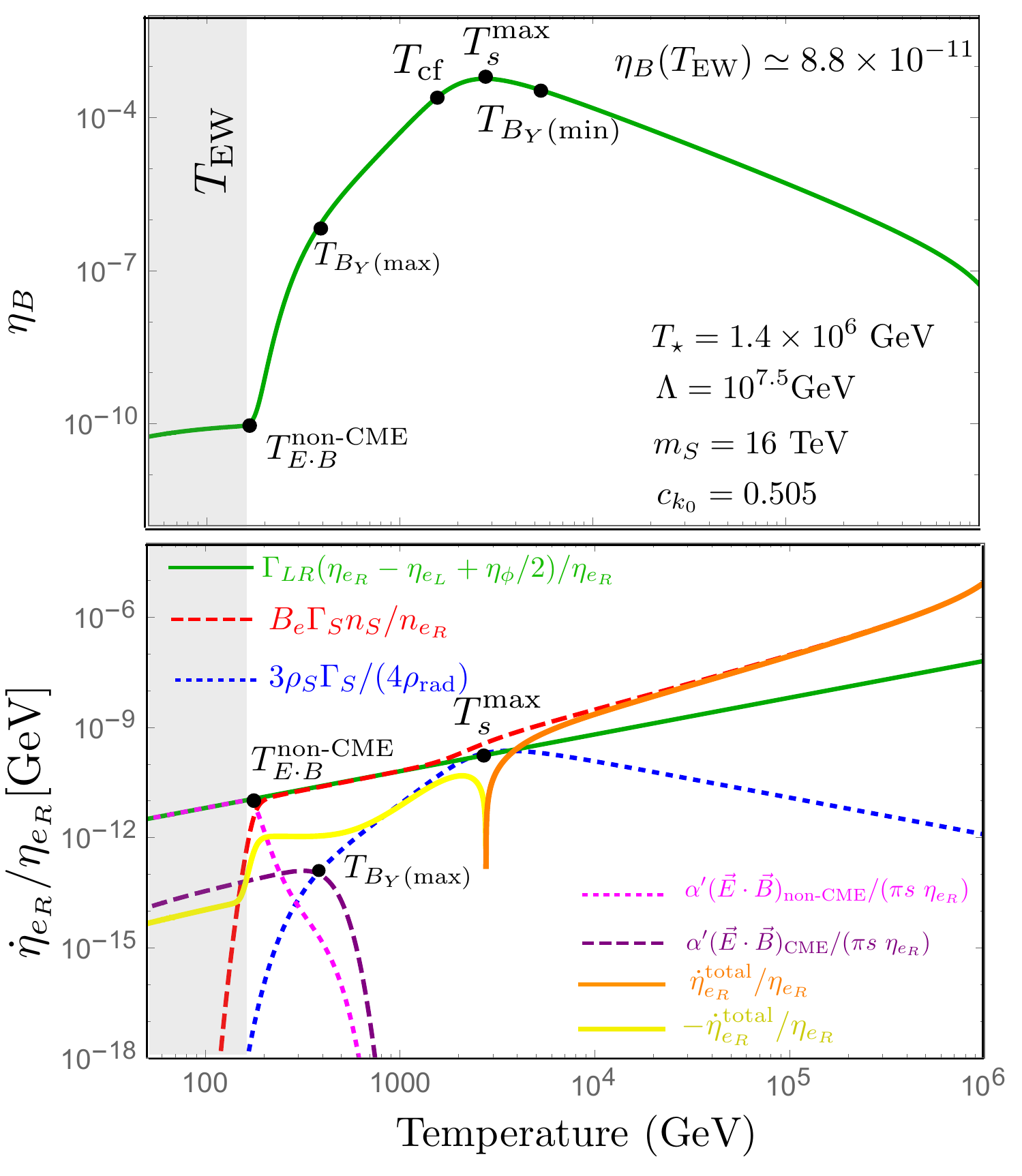}
\caption{ The evolution of $\eta_B$ for the benchmark~\ref{eq:benchmark} as a function of temperature is presented in the upper panel. The lower panel is each of the contributions to $\dot \eta_{e_R}$ (Eq.~\ref{eq:etaevol}) normalized by$\eta_{e_R}$, the quantity which is equal to $\dot \eta_B/\eta_B $ as well. In the lower panel, the solid green line is proportional to the rate of the chirality flip of the right-handed electron. The dashed red line is the relative growth rate of the asymmetry in the right-handed electron injected by the flavon. The dotted blue line is the contribution of the flavon to the dilution of the asymmetry. The dotted magenta and the dashed purple line, respectively, show the non-CME and the CME components of the hypermagnetic field effect in the evolution of $\eta_B$. The solid orange and yellow lines, together, show the magnitude of the sum of the contributions. Among the contributions, the flavon decay and the non-CME component of $(\vec E_Y \cdot \vec B_Y)$ increase the asymmetry while the rest results in a lower asymmetry. In this benchmark, the gap between when the flavon decays exponentially and when $(\vec E_Y \cdot \vec B_Y)_{\text{non-CME}}$ dominates is small and that is one of the main reasons that the asymmetry is saved at good values. } 
\label{fig:etaB2}
\end{figure}

As mentioned earlier, the evolution of $\eta_B$  is intimately related to that of $\eta_{e_R}$: $ \eta_B = \frac{198}{481}\eta_{e_R}$. Therefore, by solving Eq.~\ref{eq:erasym}, we are practically obtaining the evolution of $\eta_B$. To discuss the physical effects important in each time interval of the evolution, the evolution of the terms contributing to $\dot \eta_{e_R}/\eta_{e_R}$ (Eq.~\ref{eq:etaevol}) and $\dot B_Y/B_Y$ (Eq.~\ref{eq:By}) are shown in Figs.~\ref{fig:By2} and~\ref{fig:etaB2}.

To accomplish this task, let us first rewrite Eq.~\ref{eq:By} as 
\beq 
\frac{\dot B_Y}{B_Y} =  - 2 H - \frac{k'}{ B_Y^2} (\vec E_Y \cdot \vec B_Y)\nonumber,
\eeq
 where $\vec E_Y \cdot \vec B_Y$ in Eq.~\ref{eq:EdotB} can also be separated as 
\begin{align*}
&(\vec E_Y \cdot \vec B_Y)_{\text{non-CME}} \equiv \frac{B_Y^2}{\sigma} k',\\
&(\vec E_Y \cdot \vec B_Y)_{\text{CME}} \equiv \frac{B_Y^2}{\sigma} \cdot  \frac{\alpha'}{2\pi} c'_E.
\end{align*}
Note that, here, $c'_E$ is a negative quantity, therefore the effect of $(\vec E_Y \cdot \vec B_Y)_{\text{CME}}$ and $(\vec E_Y \cdot \vec B_Y)_{\text{non-CME}}$ are opposite of each other. 

Similarly, let us rewrite Eq.~\ref{eq:erasym} in terms of $\eta_{e_R}$:
\begin{align} 
\dot \eta_{e_R}  &+  \frac{3}{4} \frac{\rho_S}{\rho_{\text{rad}}} \Gamma_S \eta_{e_R} \nonumber\\
& = - \Gamma_{LR} \left( \eta_{e_R} - \eta_{e_L} + \frac{\eta_{\phi}}{2} \right)\nonumber\\
& + B_e \Gamma_S \frac{n_S}{s} + \frac{\alpha'}{ \pi s } \vec E_Y \cdot \vec B_Y,
\label{eq:etaevol}
\end{align}
where the derivation of this equation is presented in Appendix~\ref{app:evolution}. Notice that the second term comes from the domination of the flavon after $\T$. 
 
 According to Figs.~\ref{fig:By2} and ~\ref{fig:etaB2} , the following critical temperatures 
 can be distinguished: 
 \begin{itemize}
 \item $\TBmin:$ This is the temperature at which the HMFA is at its minimum. 
\item$\TBmax:$ This is when the HMFA reaches its maximum. 
\item$T_s^{\text{max}}$: this is when the deviation of the Hubble rate from the  SC Hubble rate is maximum. 
\item $\Tx$: As explained earlier (Fig.~\ref{fig:rates}), this is the temperature at which the chirality flip rate of the electrons equals the Hubble rate. 
\item $\Tsat$: This is the temperature at which the baryonic asymmetry saturates and becomes flat. As can be seen from Fig.~\ref{fig:etaB2}, this is the temperature at which the non-CME component of Eq.~\ref{eq:erasym} becomes the dominant effect that leads to an increase in the asymmetry.  
 \end{itemize}
\vspace{1 in} 
 {\bf The Evolution of $B_Y(T)$:}\\
 
\indent Now that we have identified the critical temperatures, we can move on to discussing the following intervals of temperature, which are identified in Fig.~\ref{fig:By2}:
 \begin{itemize}
\item [--] $ \TBmin < T < \T$: In this interval, the Hubble rate is higher than $\frac{k'}{B_Y^2} (E_Y\cdot B_Y)\simeq \frac{k'}{B_Y^2} (E_Y\cdot B_Y)_{\text{CME}}$, as illustrated in the $\dot B_Y/B_Y$ plot. This leads to a decrease of $B_Y(T)$ according to the expansion of the Universe.
\item[--] $ \TBmax < T < \TBmin$: Here, the HMFA increases rapidly due to the domination of the $\frac{k'}{B_Y^2} (E_Y\cdot B_Y)_{\text{CME}}$ over the Hubble rate.  As emphasized earlier, this is the term that makes the growth of the HMFA possible. 
\item[--] $ \TEW <T < \TBmax$: In this interval, the Hubble rate dominates, which once again leads to the decrease of the HMFA according to the expansion of the Universe.
\end{itemize}

 {\bf  The Evolution of $\eta_B(T)$:}\\
 \indent Similarly, to better comprehend the evolution of the asymmetries, let us study the plots shown in Fig.~\ref{fig:etaB2}. The upper panel is $\eta_B$ and the lower panel is the magnitude of each of the contributions to $\frac{\dot \eta_B} {\eta_B} = \frac{\dot \eta_{e_R}}{\eta_{e_R}},$ as a function of temperature.
 
In the lower panel, the solid green line is proportional to the rate of the chirality flip of the right-handed electron, which leads to the wash out of the asymmetry by the weak sphalerons. Notice that due to our choice of $y-axis$, this term is independent of $\eta_B$. The dashed red line is the relative growth rate of the asymmetry in the right-handed electron coming from the flavon. If $\eta_{e_R}$ is leaning toward zero, this term becomes greater and prevents the asymmetry from depleting. The dotted blue line represents the term that appears due to the domination of $\rho_S$ (and the decay of the flavon to radiation) at high temperatures.\footnote{It is worth mentioning that in the SC, this term does not appear.} This term is also independent of $\eta_B$ and, we will refer to it as the dilution term.

As discussed earlier, the term coming from the Abelian anomaly has two contributions: $(\vec E_Y \cdot \vec B_Y)_{\text{CME}}$ (the dashed purple line) which eats up part of the asymmetry to amplify the HMFA, and $(\vec E_Y \cdot \vec B_Y)_{\text{non-CME}}$ (the dotted magenta line) which leads to an increase in the asymmetry. The CME component is independent of $\eta_B$, but the non-CME component is proportional to the inverse of $\eta_B$. Hence, we see that the terms leading to an increase in the asymmetry are sensitive to $\eta_B$ and they grow if $\eta_B \to 0$. This is a reassurance that the system wants to save the asymmetry as much as possible.\footnote{It is clear that the physics would not change if we had plotted $\dot \eta_B$ instead of $\dot \eta_B/\eta_B$.} Finally, the solid orange and the yellow line represents the sum and the negative sum of all of these contributions. In the following, we discuss the main players in each of the temperature intervals.

 \begin{itemize} 
\item[--]  \noindent  $T_s^{\text{max}} \lesssim T< \T$: 
Here, the evolution of $\eta_B$ is mostly governed by the flavon, the effect of which is two-fold: the production of asymmetry due to the decay of the flavon, and the dilution of the asymmetry due to its effect on the expansion of the Universe (dashed blue line). As we reach $\Tx$, the chirality flip of the electron becomes relevant as well, slowing down the increase in the asymmetry. Notice that at $T_s^{\text{max}}$, there is a cancellation between the terms that increase the asymmetry and those that lead to the reduction of the asymmetry. This feature is consistent among all of the benchmarks that yield the desired values of $\eta_B$ and $B_Y$ (Eq.~\ref{eq:desired}). Thus, the asymmetry is increasing up until $T_s^{\text{max}}$, and after that starts decaying. Comparing Fig.~\ref{fig:By2} and Fig.~\ref{fig:etaB2}, we see that once the asymmetry becomes greater than $ \eta_B \gtrsim 2 \times 10^{-4}$, the HMFA starts increasing, and thus $ \TBmax <  T_s^{\text{max}} < \TBmin$. 

\item[--] \noindent $\Tsat \lesssim T \lesssim T_s^{\text{max}}$: During this interval, the rate of chirality flip of the right-handed electron (or equivalently, the rate of the washout of the asymmetry due to the sphalerons) exceeds\footnote{To be more exact, the dilution term (dotted blue line) is also important in decreasing the asymmetry. This is especially true for temperatures closer to $ T_s^{\text{max}}$. However, this term quickly drops and its effect becomes negligible at lower temperatures.} the production rate of asymmetry through the flavon decay. As a result, the asymmetry decreases. Nonetheless, as can be seen from Fig.~\ref{fig:etaB2}, these two rates are almost compatible, preventing the asymmetry from diminishing too quickly. This is an example of how the \textit{gradual} decay of the flavon to right-handed electron helps to retain the asymmetry in the fermions. 

 \item[--] \noindent $ \TEW <T < \Tsat$: As we reach $\Tsat$, the non-CME component of  $\vec E_Y \cdot \vec B_Y$ becomes compatible with the rate of electron chirality flip, which slows down the decrease of the asymmetry significantly. In other words, the amplified hypermagnetic field feeds back to the asymmetry and helps to preserve the asymmetry. Therefore, during this interval, the asymmetry is almost constant. In general, a successful benchmark is the one that there is not a large gap between $\Tsat$ and the temperature at which the flavon decays exponentially. If this gap is large, the sphalerons have enough time to wash out the asymmetry quickly.

 \end{itemize}

Now that we have discussed each of the important intervals, let us scan through the parameter space and indicate the sweet regions that give the desired values at $\TEW$ (Eq.~\ref{eq:desired}). Before that, however, allow us to emphasize two features of this benchmark that made it desirable: 1) For (most of) the temperatures bellow $T_s^{\text{max}}$, the terms leading to an increase in the asymmetry are compatible with the ones that cause the asymmetry to decrease. Generally, this means that either the flavon is long-lived which then injects the asymmetry to the fermionic sector gradually and pushes $\Tx$ closer to $\TEW$ as well, and/or the gap between $\Tsat$ and the temperature at which the flavon decays exponentially is very small. 2) There is enough time for the hypermagnetic field to grow before $\TEW$ (e.g, $\TBmax > \TEW$). However, if $\TBmax$ is at very high temperatures (e.g, $\TBmax \sim  T_s^{\text{max}}$), the $(\vec E_Y \cdot \vec B_Y)$ terms govern the evolution of $\eta_B$ and the sphalerons become subdominant. Therefore, the asymmetry is restored at higher values than desired. Some of the examples of this case will be indicated in the next subsection. In the following, we show the $\eta_B(\TEW)$ and $B_Y(\TEW)$ as a function of various parameters. 

\subsection{Scanning Parameter Space}

In this section, we scan through the parameter space and find $\eta_B(\TEW)$ and $B_Y(\TEW)$ for various values of $\Lambda,~\epsilon_m,~ \T$ and $c_{k_0}$. From a few test runs, we realize that  
$\Lambda$ should live in a narrow range of $ 5 \times 10^{6} \gev \lesssim \Lambda \lesssim  5 \times 10^{9} \gev$. For $\Lambda <5 \times 10^{6} \gev$, the maximum temperature($\Tmax$) must be below $10^4 \gev$, which means that the chirality flip of the right-handed electron process is in equilibrium from the beginning of the Universe, and therefore the weak sphalerons will wash out the asymmetry in the SM fermions as soon as the flavon starts decaying. Thereby, for $\Lambda <5 \times 10^{6} \gev$, we get $\eta_B (\TEW) \ll \eta_B^{\text{obs}}$ and the hypermagnetic field does not have a chance to amplify. For $\Lambda  \gtrsim10^9 \gev$, the decay width is too large such that flavon decays too quickly. Thus, the flavon cannot help with the preservation of the asymmetry in the early universe. In other words, $\Tx \gg \TEW$ and the weak sphalerons wash out the asymmetry.

\begin{figure}[h!]
\includegraphics[width=0.5\textwidth, height=0.43\textheight]{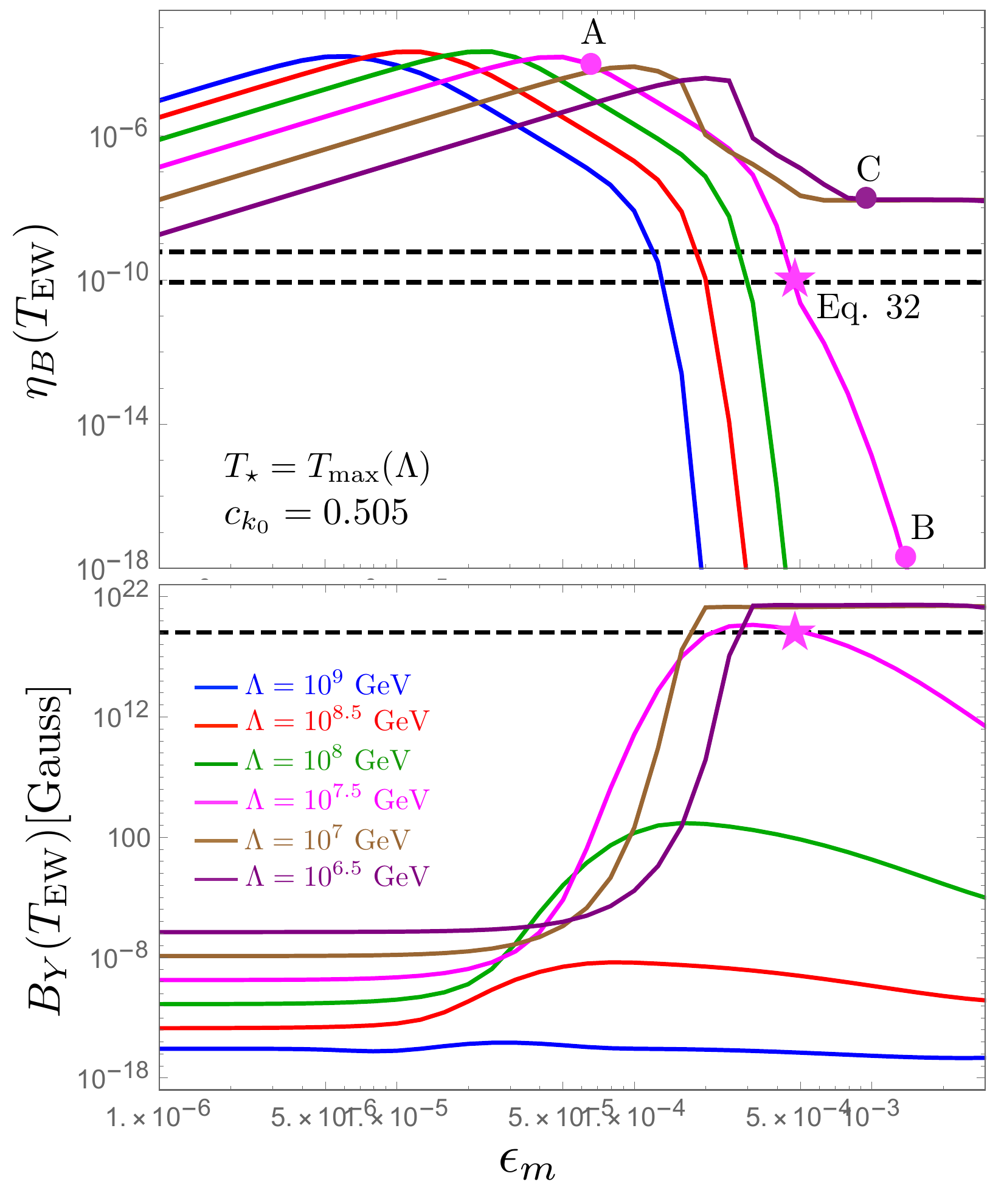}
\caption{The dependence of $\eta_B$(upper panel) and hypermagnetic field amplitude $B_Y$ (lower panel) on the mass of the flavon $m_S$. The solid lines represent different cut-off scales $\Lambda$.} 
\label{fig:etaByLambda}
\end{figure}

Fig.~\ref{fig:etaByLambda} presents the baryon asymmetry of the Universe and the HMFA at $T= \TEW$ as a function of $\epsilon_m$. Different curves represent different values of $\Lambda$, and we have fixed $\T =\Tmax (\Lambda)$ and $c_{k_0} = 0.505$. For a fixed $\Lambda$, if $\epsilon_m$ is very  small ($\ll 10^{3.5} \text{GeV}/\Lambda$ or equivalently $m_S \ll 3 \TeV$-- Benchmark A), the injection of asymmetry from the flavon to right-handed electron occurs at a slow rate. This effect has two consequences; 1) we move away from SC so much that $\Tx$ is after $\TEW$, and 2) the asymmetry is not large enough to amplify the HMFA. Since $\Tx < \TEW$, the value of $\eta_B(\TEW)$ will just depend on the work of the flavon. Therefore, in this regime, as we increase $\epsilon_m$ we see that $\eta_B(\TEW)$ is increasing because more of the flavon has decayed into the right-handed electron. An example of such a case is studied in Appendix~\ref{app:badbench}. 

On the other extreme, for large $\epsilon_m$, the flavon decays too quickly\footnote{ Note that $\Gamma_S$ is proportional to $\epsilon_m^3 \Lambda$, and thus as we increase either of $\epsilon_m$ or $\Lambda$, the decay width of the flavon increases and the flavon decays faster. However, the injection of asymmetry to the right-handed electron is proportional to $\Gamma_S/ m_S = \epsilon_m$, and an increase in $\epsilon_m$ causes more asymmetry to be transferred to the right-handed electron. Thereby, for large $\epsilon_m$, depending on the value of $\Lambda$, we either end up with too much asymmetry (Benchmark C) or too little asymmetry (Benchmark B).} . In this scenario, the value of $\eta_B(\TEW)$ is intimately connected to the HMFA. If $B_Y$ has not been amplified, the effect of $(\vec E_Y \cdot \vec B_Y)$ is negligible and the sphalerons have enough time to eat up the asymmetry. Hence, in such cases, we see that both $\eta_B(\TEW)$ and $B_Y(\TEW)$ are small (e.g, Benchmark B). If $B_Y$ has been amplified, the battle between $(\vec E_Y \cdot \vec B_Y)_{\text{non-CME}} +$ flavon decay vs. $(\vec E_Y \cdot \vec B_Y)_{\text{CME}}$ determines the evolution of $\eta_B$ and the effect of the chirality flip of right-handed electrons is negligible (e.g, Benchmark C). Let us mention that benchmark C will be acceptable if we relax the assumption that the value of $\eta_B$ and $B_Y$ stay fixed during and after the EWPT. 

For some values of $\Lambda$, the intermediate values of $\epsilon_m$ yield Eq.~\ref{eq:desired}. That comes from a delicate work of non-standard cosmology (domination of $\rho_S$), and the effect of $(\vec E_Y \cdot \vec B_Y)_{\text{non-CME}}$ at lower temperatures. Both of these effects are important in taming the work of sphalerons, as explained in Section.~\ref{sec:case}. 

\begin{figure}[h!]
\includegraphics[width=0.5\textwidth, height=0.42\textheight]{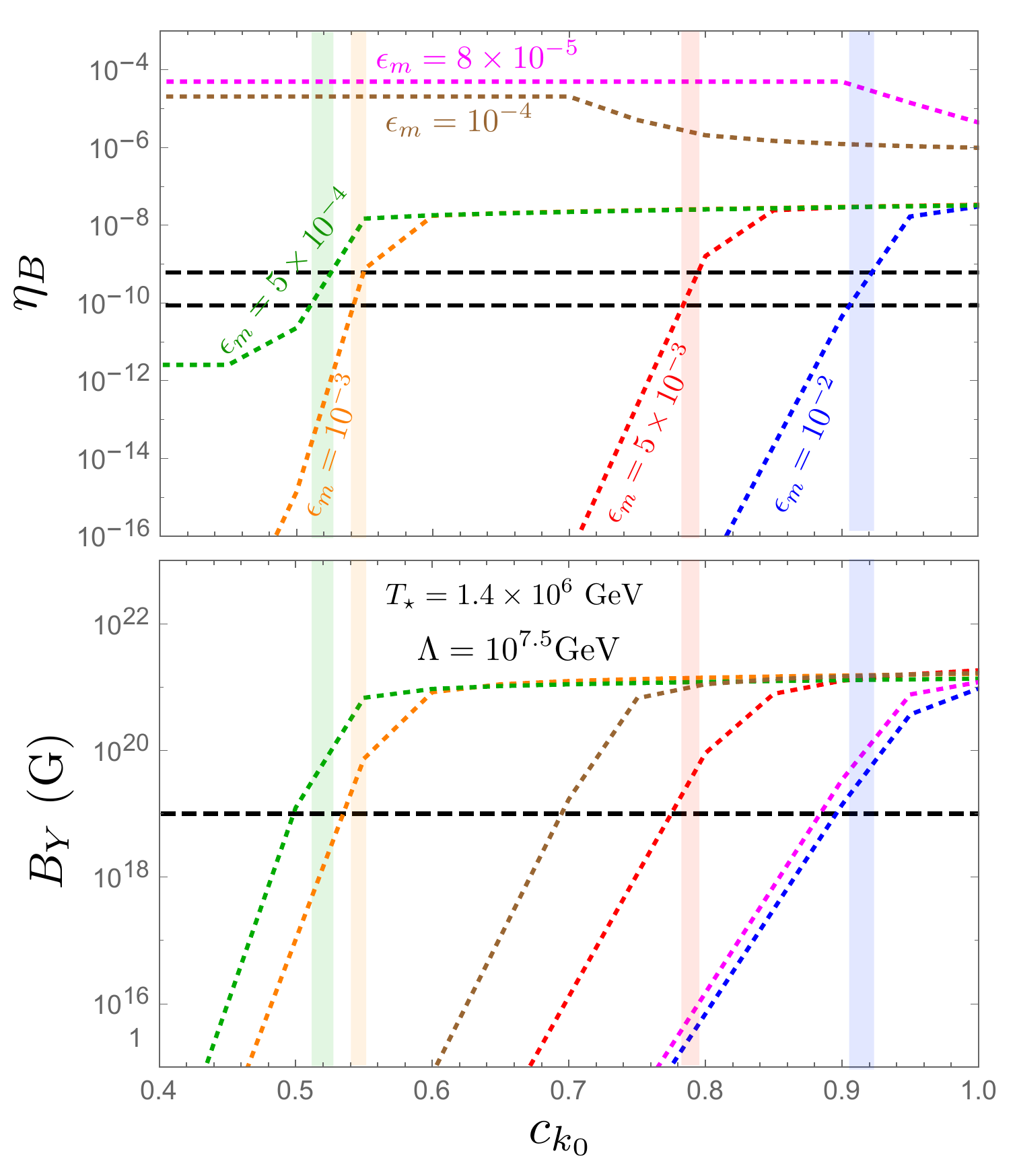}
\caption{The dependence of $\eta_B$(upper panel) and hypermagnetic field amplitude $B_Y$ (lower panel) on $c_{k_0}$. Each of the dashed lines represent a fixed $\epsilon_m$, and for all of the curves, we have fixed $\Lambda = 10^{7.5} \ \gev$. The results are highly sensitive to the value of $c_{k_0}$ and in general the closer we get to $c_{k_0} =1$, the higher the values of $\eta_B$ and $B_Y$ become. Near $c_{k_0} =1$, we see a stabilization for all of the values of $\epsilon_m$ we considered. For $\epsilon_m \lesssim 10^{-4}$ we see that we consistently get a large value of $\eta_B$, while $B_Y$ exponentially decreases to zero.}
\label{fig:ck0}
\end{figure}

Another free parameter that can affect the baryon asymmetry and especially hypermagnetic field is $c_{k_0}$. As the plots in Fig.~\ref{fig:ck0} show, the results are sensitive to $c_{k_0}$ and we cannot choose an arbitrary small $c_{k_0}$ value. Since $c_{k_0}\leq1$, it mainly affects the HMFA through the CME contribution to its evolution equation. Interestingly, the value of $c_{k_0}$ is particularly important in the non-CME contribution to the evolution of the asymmetry, whose main role is to save the asymmetry at temperatures closer to $\TEW$. Therefore, as illustrated in Fig.~\ref{fig:ck0}, large values of $c_{k_0}$ will result in large $\eta_B(\TEW)$ and $B_Y(\TEW)$. Similarly, for small values of $c_{k_0}$, the opposite is true. However, if $\epsilon_m$ is very small, the evolution of $\eta_B$ is only determined by the flavon dynamics (see Appendix.~\ref{app:badbench}) and it is independent of $(\vec E_Y\cdot \vec B_Y)$. In these cases (e.g, $\epsilon_m= 8 \times 10^{-5}, 10^{-4}$ in Fig.~\ref{fig:ck0}), the baryonic asymmetry at $T= \TEW$ will remain large even for small $c_{k_0}$. The desirable values of $c_{k_0}$ for each $\epsilon_m$ is presented by a shaded band in Fig.~\ref{fig:ck0}.

Yet, one other parameter that can leave an impact on the final result is the value of $\T$. We have discussed the maximum value of $\T$, but technically we can choose any $\T$ less than $\Tmax$. By examining different values of $\T$, we noticed that only $\T = \Tmax$ gives the best results. If $ \T \ll \Tmax$, the flavon does not have enough time to efficiently transfer its asymmetry to the fermionic sector and eventually leads to the amplification of the HMFA. 

 %*********************EVOLUTION*********************** 
\section{ Conclusion }
\label{sec:conclusion}
%*********************EVOLUTION*********************** 

In this paper, we discussed the possibility of the simultaneous generation of baryonic asymmetry to $\eta_B (\TEW) \sim 8.5 \times 10^{-11}$ and the amplification of HMFA from a small seed to $B_Y (\TEW) \sim 10^{20}\G$ in the presence of a flavon that carries a large asymmetry $(\xi_S =1)$. We found a successful scenario that lives in a region where the cutoff scale is $\sim10^{7.5}\gev$. Given the cutoff scale, the mass of the flavon could vary over a small range of values to give a desirable outcome. Another free parameter that played an important role in the dynamics of baryonic asymmetry and HMFA was the comoving wavenumber of the hypermagnetic field.  According to our study, the comoving wavenumber should be $(0.5 -1) \times 10^{-7}$. In general, we found there is a strong sensitivity to each of these parameters. A small change could result in a drastic change in the results. This is because we need a delicate cancellation between the terms that increase the asymmetry and the ones that result in a lower asymmetry, and thus we have to choose our parameters carefully. 

For most of the parameter space, we get a large asymmetry of the Universe, while having a small (compared to the desired) value of HMFA. This occurs because the asymmetry in the flavon transfers into baryonic asymmetry, but there is not enough time for the HMFA to grow. This occurs in the benchmarks where the flavon is very long-lived and it transfers its asymmetry to the fermionic sector at a very small pace. The baryonic asymmetry must reach above $\eta_B > 2 \times 10^{-4}$ for the HMFA to start growing. When the flavon decays too slowly, the baryonic asymmetry reaches $2 \times 10^{-4}$ either very close or even after the electroweak temperature. Therefore, the HMFA remains small.

On the other hand, if the flavon is very short-lived, we may have two very different cases depending on the value of $\Lambda$: 1) For sufficiently small $\Lambda$, the flavon injects its asymmetry to fermionic sector quickly and causes the HMFA to grow fast. The hypermagnetic field then feeds back to the asymmetry and prevents it from being washed out. In such scenarios, we noticed that we end up with a larger $\eta_B$ than expected. 2) for a relatively bigger $\Lambda$, we may also have a case where the injection of asymmetry is inefficient and we end up with a very small $\eta_B$. 

The value of the comoving wavenumber has an indisputable effect on the HMFA and thus has a great influence on the baryonic asymmetry as well. We noticed that to get the observed value of $\eta_B$ and the desired value of $B_Y(\TEW)$, we have to live in a small region of the parameter space and thus our scenario is predictive. 

In this paper, we only considered the coupling of the flavon with the first generation of fermions. This choice was suitable because the branching ratio of the flavon to electrons was enhanced. In fact, we could not find a benchmark that could explain both baryogenesis and magnetogenesis with other choices of flavon coupling. Having said that, the assumption of flavon coupling to only the first generation of fermions is theoretically justifiable as well. That is because the first generation of fermions is much lighter than the electroweak scale. Thus, explaining their small Yukawa couplings is of priority.

 %*********************EVOLUTION*********************** 
\acknowledgments 
%*********************EVOLUTION*********************** 
We would like to thank S. Ipek, J. Kopp, H. Mehrabpour, M. Shaposhnikov, and G. White for numerous useful conversations. We are also thankful to the CERN theory division and Mainz Cluster of Excellence for their hospitality. We also would like to express our gratitude to B. Enshaeian for his continuous support. 

\bibliography{genesis}
\newpage
\appendix
 
 %*********************EVOLUTION*********************** 
\section{ Evolution of $\eta_R$}
\label{app:evolution}
%*********************EVOLUTION*********************** 
The asymmetry in the number density of the right-handed electrons is governed by the following Boltzman equation:
\begin{align} 
\dot n_{e_R} &+ 3 H n_{e_R}  = - \Gamma_{LR} \left( n_{e_R} - n_{e_L} + \frac{n_{\phi}}{2} \right)\nonumber\\
&+ B_e \Gamma_S n_S +\frac{\alpha'}{\pi} \vec E_Y \cdot \vec B_Y. 
\end{align}
By dividing both sides by the co-moving entropy\footnote{Note that
\begin{align*}
\dot \eta &= \frac{ d(n/s)}{dt} = \frac{\dot n}{s} - \frac{n}{s^2} \frac{ds}{dT}\dot T= \frac{\dot n}{s} - 3 \frac{n}{s} \frac{\dot T}{T}.
\end{align*}}, $s$, we can convert this equation to an equation describing the evolution of $\eta_{e_R}$.
\begin{align}
\dot \eta_{e_R}  + 3 \eta_{e_R} \frac{\dot T}{T} + 3 H \eta_{e_R}& = - \Gamma_{LR} \left( \eta_{e_R} - \eta_{e_L} + \frac{\eta_{\phi}}{2} \right)\nonumber\\
&+ B_e \Gamma_S \frac{n_S}{s} + \frac{\alpha'}{ \pi s } \vec E_Y \cdot \vec B_Y
\label{eq:appeta}
\end{align}
On the other hand, by definition we know $\dot T / T = 1/4 \dot \rho_{\text{rad}}/\rho_{\text{rad}}$. This is while we can use Eq.~\ref{eq:density} to find $\dot \rho_{\text{rad}}/\rho_{\text{rad}}$:
\beq 
\frac{\dot \rho_{\text{rad}}}{\rho_{\text{rad}}}  + 4 H = \Gamma_S \frac{\rho_S}{\rho_{\text{rad}}}.
\eeq
Hence, Eq.~\ref{eq:appeta} becomes
\begin{align} 
\dot \eta_{e_R}  &+  \frac{3}{4} \frac{\rho_S}{\rho_{\text{rad}}} \Gamma_S \eta_{e_R} \nonumber\\
& = - \Gamma_{LR} \left( \eta_{e_R} - \eta_{e_L} + \frac{\eta_{\phi}}{2} \right)\nonumber\\
& + B_e \Gamma_S \frac{n_S}{s} + \frac{\alpha'}{ \pi s } \vec E_Y \cdot \vec B_Y
\end{align}
In the SC, $\rho_S \simeq0$, and thus the second term is negligible. In our case, however, the second term becomes important in some interval of the temperature. 
 
 %*********************EVOLUTION*********************** 
\section{ Bad Benchmarks}
\label{app:badbench}
%*********************EVOLUTION*********************** 

Here we show the $\eta_B$ evolution for the benchmarks shown in Fig.~\ref{fig:etaByLambda}. 
 \begin{figure}[H]
\includegraphics[width=0.52\textwidth, height=0.44\textheight]{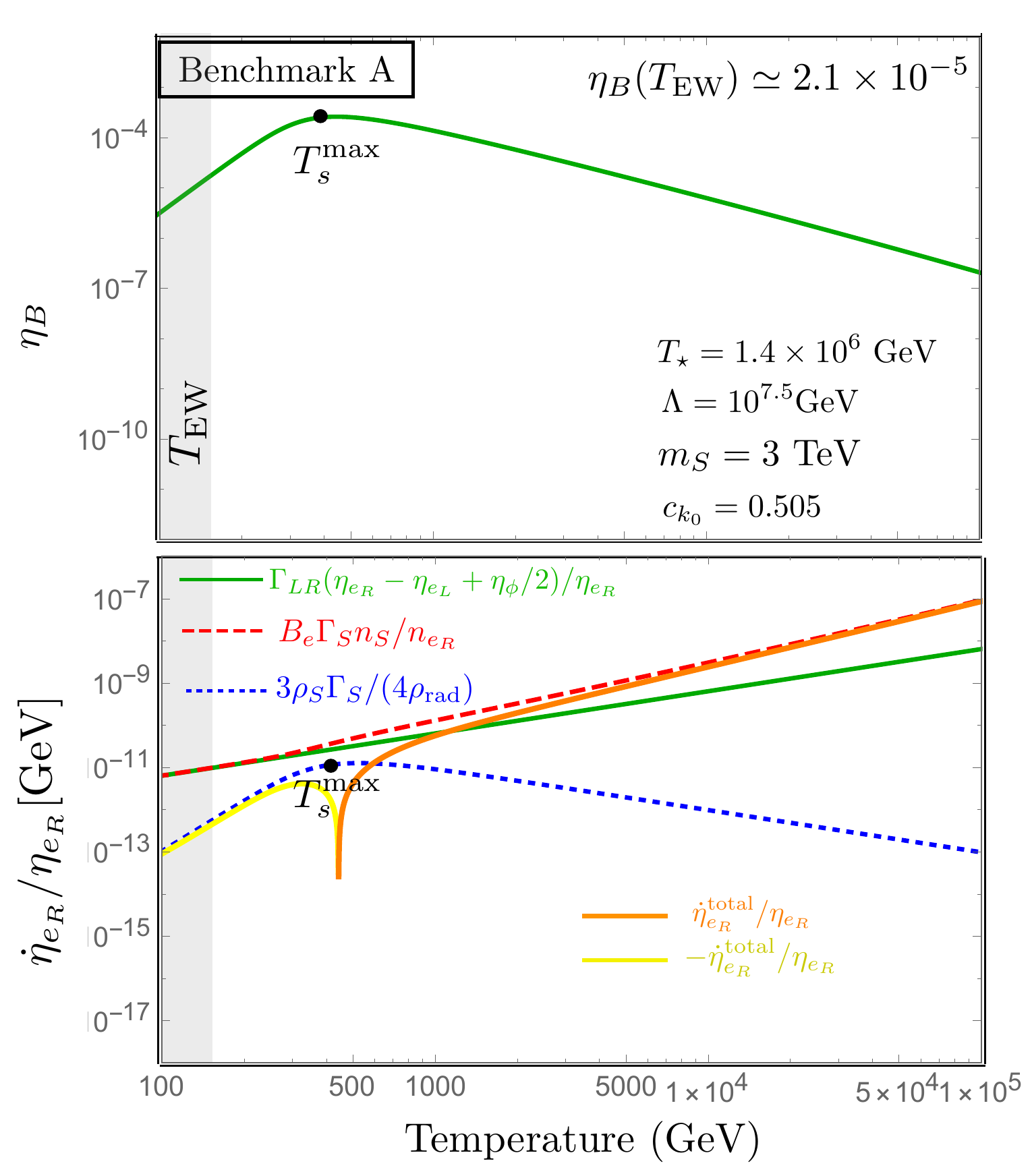}
\caption{ The evolution of $\eta_B$ for benchmark A as a function of temperature is presented. The lower panel is each of the contributions to $\dot \eta_{e_R}$ (Eq.~\ref{eq:etaevol}) normalized by $\eta_{e_R}$, the quantity which is also equal to $\frac{\dot \eta_B}{\eta_B} $. In the lower panel, the solid green line is proportional to the rate of the chirality flip of the right-handed electron. The dashed red line is the relative growth rate of the asymmetry in the right-handed electron injected by the flavon. The dotted blue line is the contribution of the flavon to the dilution of the asymmetry. The solid orange and yellow lines, together, show the magnitude of the sum of the contributions. In this benchmark, the asymmetry is slowly transferring from the flavon to the right-handed electron. This rate is very small, which means the rate of flavon depletion is small. Hence the cosmology becomes very non-standard. In this case, $\Tx$  occurs too close to $\TEW$, and thus the asymmetry is not washed out efficiently. Hence, we end up with too large $\eta_B(\TEW)$. The HMFA does not have a chance to grow and stays close to its initial value.  } 
\label{fig:benchA}
\end{figure}
 \begin{figure}[H]
\includegraphics[width=0.52\textwidth, height=0.44\textheight]{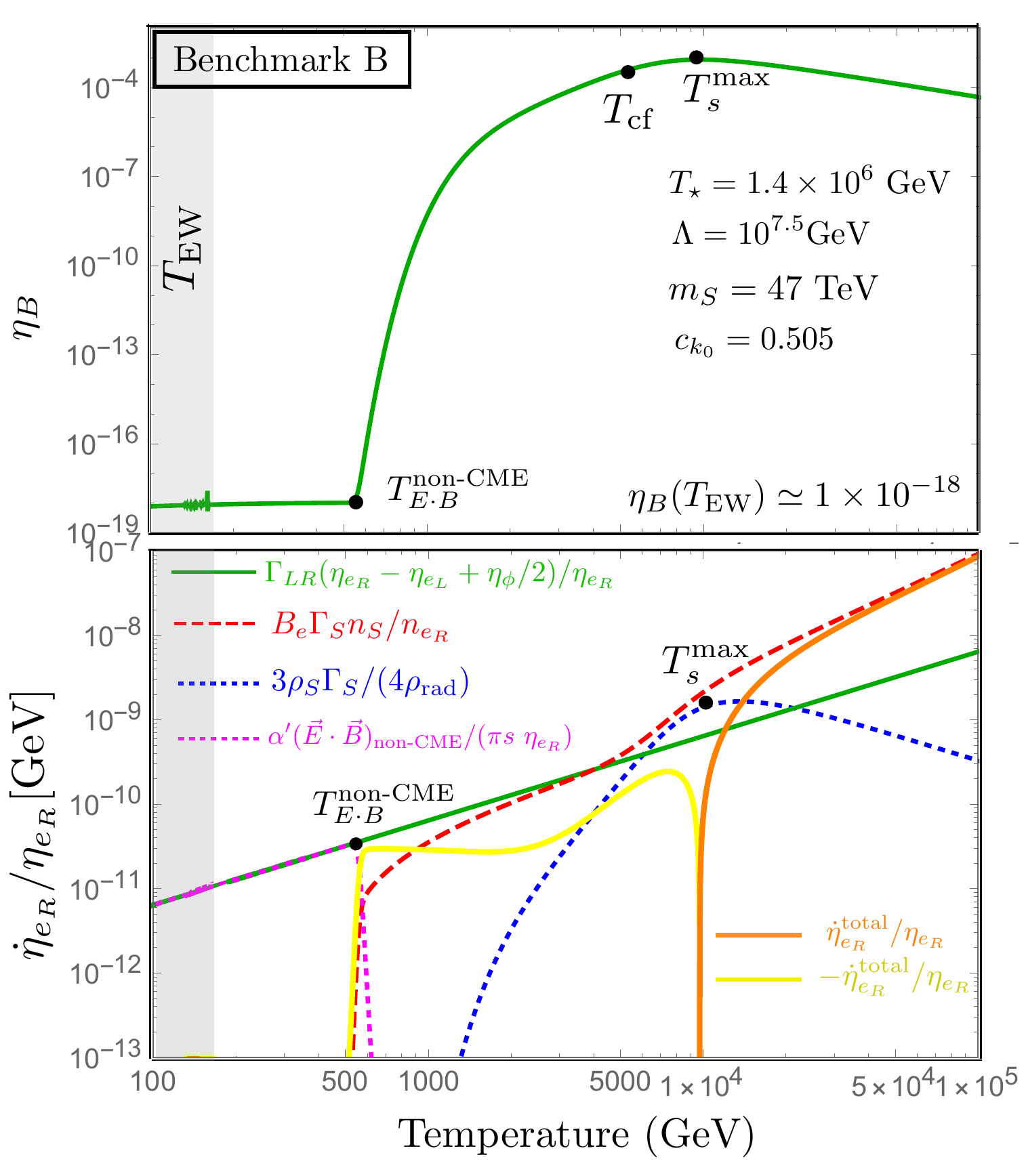}
\caption{ The evolution of $\eta_B$ for benchmark B as a function of temperature is presented. The lower panel is each of the contributions to $\dot \eta_{e_R}$ (Eq.~\ref{eq:etaevol}) normalized by $\eta_{e_R}$, the quantity which is also equal to $\frac{\dot \eta_B}{\eta_B} $. In the lower panel, the solid green line is proportional to the rate of the chirality flip of the right-handed electron. The dashed red line is the relative growth rate of the asymmetry in the right-handed electron injected by the flavon. The dotted blue line is the contribution of the flavon to the dilution of the asymmetry. The dotted magenta shows the non-CME component of the hypermagnetic field effect in the evolution of $\eta_B$. The solid orange and yellow lines, together, show the magnitude of the sum of the contributions. In this benchmark, there is a large gap between when the flavon decays and $(\vec E_Y \cdot \vec B_Y)_{\text{non-CME}}$ dominates. During this gap, the sphaleron has enough time to eat up the asymmetry and thus we are left with too little asymmetry. Since the asymmetry depletes very quickly, the CME cannot increase the HMFA efficiently.   } 
\label{fig:benchB}
\end{figure} 
\begin{figure}[h!]
\includegraphics[width=0.52\textwidth, height=0.44\textheight]{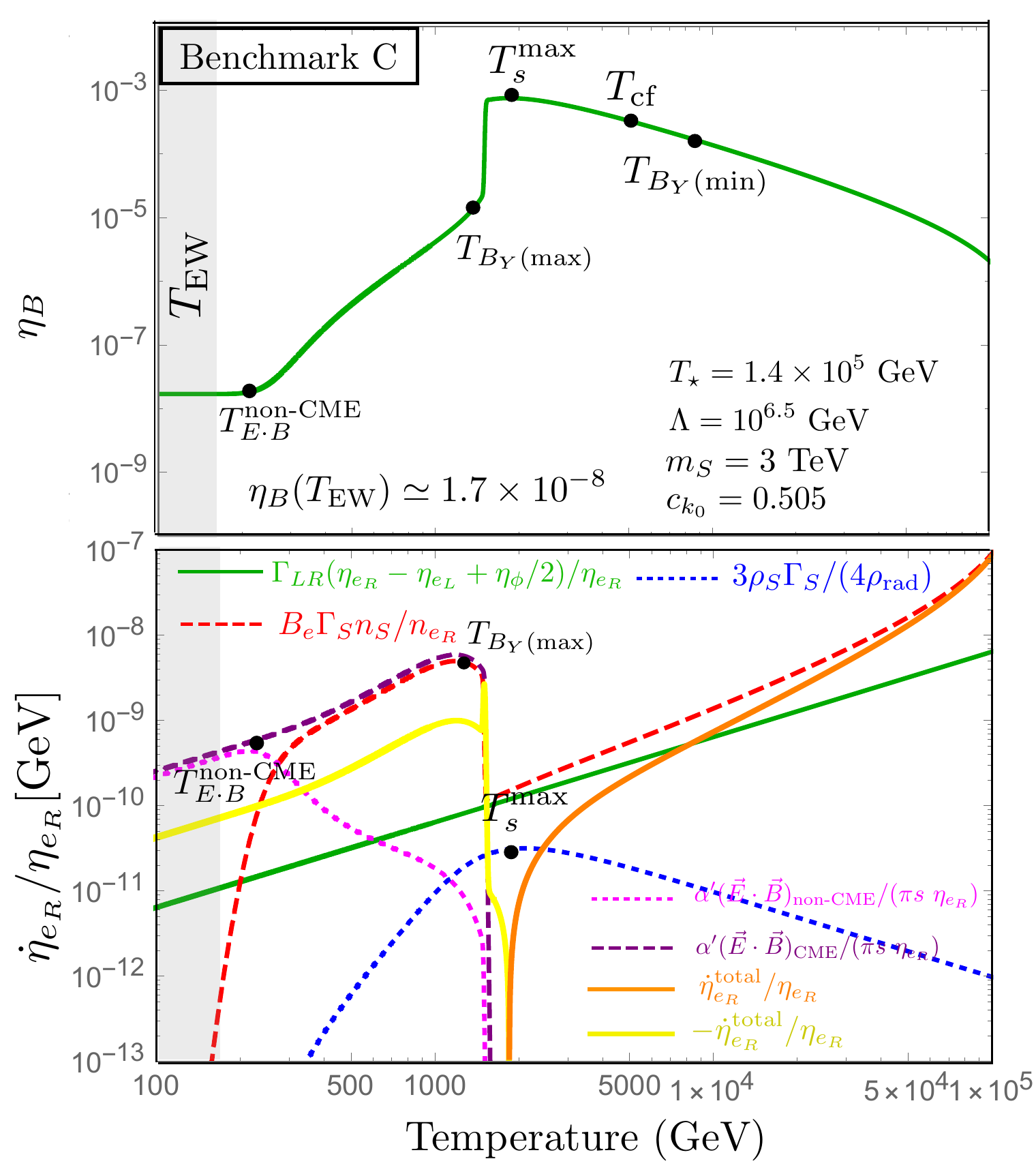}
\caption{ The evolution of $\eta_B$ for benchmark C as a function of temperature is presented. The lower panel is each of the contributions to $\dot \eta_{e_R}$ (Eq.~\ref{eq:etaevol}) normalized by $\eta_{e_R}$, the quantity which is also equal to $\frac{\dot \eta_B}{\eta_B} $. In the lower panel, the solid green line is proportional to the rate of the chirality flip of the right-handed electron. The dashed red line is the relative growth rate of the asymmetry in the right-handed electron injected by the flavon.  The dotted blue line is the contribution of the flavon to the dilution of the asymmetry. The dotted magenta and the dashed purple line, respectively, show the non-CME and the CME components of the hypermagnetic field effect in the evolution of $\eta_B$. The solid orange and yellow lines, together, show the magnitude of the sum of the contributions. In this benchmark, the flavon decays quickly and thus increases the asymmetry in $e_R$ at relatively high temperatures. Due to the increase in the asymmetry, the HMFA has enough time to amplify and even dominate the effect of the sphaleron. We see a sharp drop in the asymmetry at $\TBmax$ followed by another decline. Both of these effects are due to the CME. Eventually, the non-CME component dominates and preserves the asymmetry. However, the value at which it preserves the asymmetry is larger than the observed value. Notice that the sharp increase in the flavon term is because of the normalization.  }
\label{fig:benchC}
\end{figure}

%\bibitem{Sakharov} 
%A. D. Sakharov, \textit{Violation of CP Invariance, c Asymmetry, and Baryon Asymmetry of the Universe}, Pisma Zh. Eksp. Teor. Fiz. \textbf{5} (1967) 32 [JETP Lett. \textbf{5} (1967) 24] [Sov. Phys. Usp. \textbf{34} (1991) 392] [Usp. Fiz. Nauk \textbf{161} (1991) 61].

%\bibliographystyle{JHEP}
\end{document}